Banner appropriate to article type will appear here in typeset article

# Spontaneous Axis-Switching Swirling in Transonic Core-Annular Flow of Highly Differentiated Fluids

**Wayne Strasser[1] †, Aleck Tilbrook[2] and Daniel Wilson[3]**

[1]Department of Mechanical Engineering, Liberty University, Virginia, USA

[2]Los Alamos National Laboratory, New Mexico, USA

[3]Department of Mechanical Engineering, Clemson University, South Carolina, USA

Forcing fluids to interact at high speeds creates a multitude of instabilities. Prior studies considered the consequences of these instabilities for core-annular flow (CAF) or atomization systems. However, the existence of spontaneous axis-switching swirling (SASS) modalities in CAF have never been explored. We computationally demonstrate this new behavior for a CAF of steam surrounded by a cold non-Newtonian fuel. Wave pile-up created 3-D fuel surface disturbances which eventually became thin slender lobes with radial modal deformation. Those slender lobes were subject to further instabilities and resulted in oscillatory tangential velocities that were a large fraction of the core steam velocity and orthogonal to the axially flowing base fuel wave. Since these tangential velocities were not externally imposed by atomizer geometry or boundary conditions, it was concluded that they are self-generating. Frequency analysis revealed that SASS tended to operate at the base wave frequency. We additionally explored numerical uncertainty with our previously validated computational tool and found that SASS behavior can be achieved on a coarse, limited mesh at relatively low simulation times.

## 1. Introduction

At the intersection of Kelvin Helmholtz instabilities (KHI), Rayleigh-Taylor instabilities (RTI), Rayleigh-Plateau instabilities, Taylor-Culick flow, multiphase flow, and 3-D turbulence lives primary air-blast atomization (Oshima & Sou (2024)). KHI and RTI normally cooperate as is discussed in Jiang & Ling (2021), where KHI act in the streamwise direction and RTI act in the transverse direction (azimuthal in a cylindrical system). For atomization, a contiguous liquid fuel is violently disintegrated into fuel satellites of spatiotemporally variable shapes and sizes by a co-flowing high-speed gas phase. Typically, the two phases' interaction is facilitated by a purposed device containing an orifice (Strasser (2019)). Primary atomization is fundamentally unsteady and, due to sharp viscosity and density gradients *within* individual computational mesh elements, is costly to replicate computationally. Many have endeavored to do so, and it would be impossible to provide an exhaustive reference list for those. While atomization is prevalent in devices specifically designed so, it is also present in large-body waves under the influence of high-speed winds or breaking waves (Mostert & Deike (2020)).

† Email address for correspondence: wstrasser@liberty.edu

Although the majority of atomizers involve a centrally fed cylindrical fuel phase, prior research has been carried out on a particular style of air-blast atomization in which a central cylindrical steam flow is passed transonically over a non-Newtonian radially fed slurry fuel flow. The hot steam heats the fuel, reducing its viscosity. The result is a periodic wave-generating atomizer which offers two stages of primary atomization. The first stage is the stripping of droplets off of the thinning wave crests which are contorted by the compressed, accelerating steam. The second stage results from the crashing of those waves into annular sheets formed by past waves. The process has been referred to as "Wave-Augmented Varicose Explosions" (WAVE) and was discussed by Strasser (2021), Wilson & Strasser (2022*b*), Wilson & Strasser (2022*a*), Wilson *et al.* (2023*a*), and Wilson *et al.* (2023*b*). The idea with the incorporation of "varicose" was that, although varicose and sinuous modes were prevelant, the dominant wave patterns leaving the atomizer were mainly varicose in character. WAVE is illustrated via an instantaneous snapshot in figure 1, where flow is generally from left to right. Gas (steam in this case) is provided via the central cylinder, while non-Newtonian slurry fuel is fed radially and becomes an annular flow with a self-determined variable thickness. The nozzle's center plane and outlet have been colored by velocity magnitude contours over the range shown. Where the steam meets fuel, steam is accelerating, and waves are being produced. As was demonstrated in Wilson & Strasser (2022*b*), Wilson & Strasser (2022*a*), Wilson *et al.* (2023*a*), the waves are pronounced and block a significant area for steam to flow, causing pressure to build on the steam inlet. The blockage and wave production are responsible for periodicity, as was documented in those articles. Key components of the wave production process are the fuel "wave pool" (larger annular area just outside of the steam) and "beach" (flat of the nozzle orifice inner surface) and were discussed therein. Also elucidated were data analysis methodology, wave character, KHI, RTI, and frequency content of various quantities at various locations. Moreover, there was discussed a secondary instability cycle caused by a temperature- and shear-dependent fuel viscosity, mesh independence, the adequacy of the use of the shear-stress transport (SST) turbulence model (more validation discussion later), and the potential for secondary atomization as assessed by the turbulent Weber number. A similar but simpler system was investigated in Liu *et al.* (2022). In their case, however, the axial KHI and circumferential RTI did not develop interior to the atomizer. Additionally, no pulsations were discussed. Atomization is known to be highly unsteady (Wu *et al.* (2024)), and outlined above is an obviously globally unsteady WAVE system.

What was not noticed until now, however, is the existence of organized and directionally switching tangential velocities within the WAVE atomizer which will be demonstrated in upcoming data analyses. In a search for this behavior's origins, modal deformations like those in figure 2 were found. This instantaneous snapshot is of Mach number contours ranged linearly from 0.0 (dark) to 0.5 (light) on a cross-cutting plane which is 0.42 orifice diameters downstream of the nozzle orifice. (The cross-cutting plane used for this figure is like that shown in 1 as the "sample plane" cross-cutting the nozzle orifice. In this paper, but not considered in our past work, we will employ many of those planes. Some are upstream of the orifice, some are across the orifice, and some are downstream. More details and animations of planar information will be provided later.) The flow is moving towards the reader, and the hot steam core is seen inside the annular distorted cold fuel wave. While this result is from an earlier, somewhat coarse and preliminary 180° simulation, it led us to look deeper into the flowfield. We began considering the cause of these radial flow deformation patterns in what is now clearly manifesting in our atomizer as core-annular flow (CAF). We aimed with this research to connect CAF to modal deformations. These radial patterns are common in other fields, and more information can be found in the Appendix. Additionally, CAF is a well-studied and broad subject. The reader is encouraged to see the Appendix for a relevant CAF synopsis which lays the context for the current paper. Also included is a large table of dimensionless numbers in Table 5 which positions our work in the broader CAF field. Specifically, the Appendix contrasts the current spontaneous axis-switching swirling (SASS) mechanism with known CAF phenomena such as corkscrew flows.

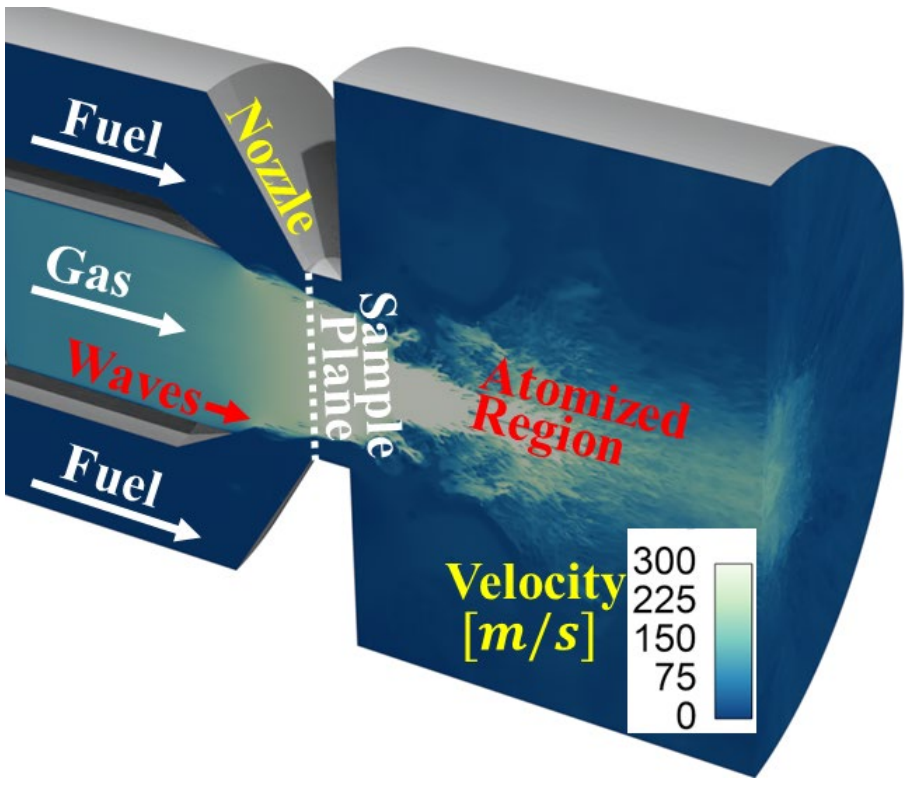

Figure 1: Instantaneous snapshot of "WAVE" atomization of non-Newtonian fuel (radially fed, becoming an annular fluid) by a hot steam core, generally flowing from left to right, pulsing at $\sim$ $1000 Hz$. While the model is a full azimuth (360°), only a half model is shown here. The atomizer center plane and exit are colored with contours of velocity magnitude, and a relevant time-lapse animation can be found here. In the animation, flow is generally from top-down, and the contours represent Mach number on the nozzle's centerplane.

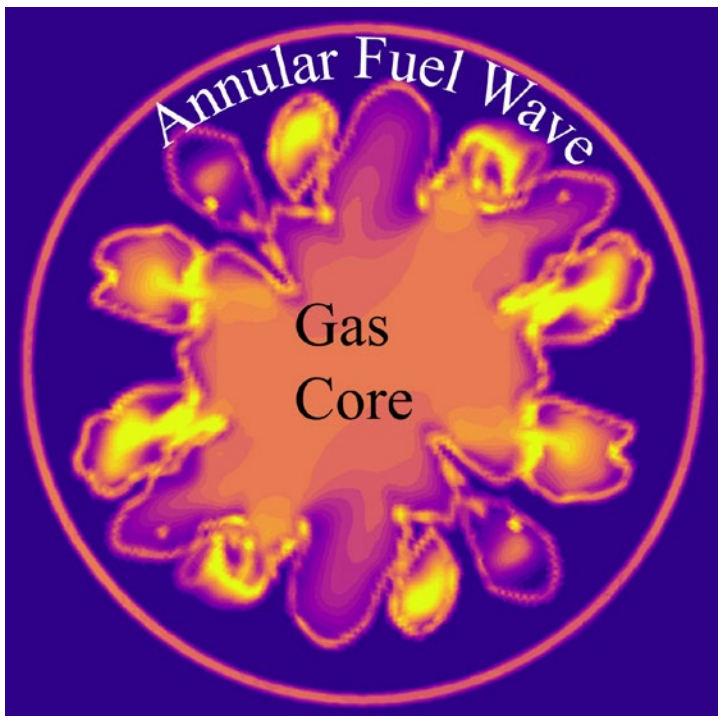

Figure 2: Instantaneous snapshot of Mach number on a cross-plane at 0.42 orifice diameters downstream from the orifice exit for a preliminary and relatively coarsely meshed 180° simulation; bulk flow is moving towards the reader and in parallel with the nozzle's axis. The color range is linear from 0.0 = dark to 0.5 = light.

## 2. Objectives

Demonstration of SASS character is sought. The swirling behavior is self-generated, as it is not imposed by geometry or boundary conditions.. Various elements missing from the current state of research leave us seeking more explanation and details. First, the WAVE atomizer was only assessed for its 3-D atomization behavior and 2-D wave formation. Nothing about self-induced directionally alternating swirl was found in the past, in terms of its existence or its cause(s). We will illustrate that the root of this swirl lies in framing the WAVE atomizer as a fully 3-D CAF. Second, the WAVE-based CAF is far different from traditional CAF. Not only is WAVE-based CAF significantly augmented by the core fluid's compressibility via a rare "block and release" mechanism described in Wilson & Strasser (2022*b*), but also the annular fluid's thickness is self-determining from the moment of its introduction. Fuel is fed into a relatively stagnant pool and *drawn out* by waves. Neither of these is the case in traditional CAF. Third, gravity is irrelevant for WAVE-CAF. Fourth, traditional CAF involves Newtonian fluids. Fifth, an "undisturbed film" never exists in the WAVE system. Sixth, few papers deal with gas-liquid density gradients. Seventh, many of the dimensionless number values shown in table 5 are well outside the range of those previously considered. Of particular interest is the rapid development of instabilities which lead to swirl; our waves are extremely "short" as determined by $L_{1,LS}$ and $L_{2,LS}$, leaving no room for wave trains. Also from table 5 is the fact that our surface tension parameters lead to very different conclusions about the relative importance of surface tension, depending on which parameter is emphasized. Eighth, few studies have dealt with surface waves blocking core flow to the extent the core flow is compressed. A ninth unique aspect of this work is the variable cyclical character of the relative core and annular flow areas. Finally, only 1 past study entailed 3-D CFD to evaluate and characterize the resulting annular liquid sheet.

Consequently, we aim to characterize and explain the root causes of a newly documented phenomenon which combines radial modal deformation with fully 3-D transonic CAF. This newfound behavior involves a rapid, direction-switching self-generated swirl of a non-Newtonian fuel with core blockage sufficient to compress the core fluid. We will demonstrate that something other than the previously discovered mechanisms for CAF-induced waves are responsible for the pulsing swirl. In short, it will be shown that RTI and KHI, running orthogonal to the primary flow direction, produce the 3-D instabilities necessary for the swirl.

## 3. Methods

### 3.1. *Governing Equations*

Following the work of many others to be cited, we use ANSYS Fluent forms of solution methodologies related to VOF (Baayoun *et al.* (2022),Terrington *et al.* (2022)), Reynolds-Averaged Navier-Stokes (RANS, Rezaeiravesh *et al.* (2023), Richards *et al.* (2023), and Chen *et al.* (2025)), and more specifically the shear-stress transport (SST) turbulence model (Anaclerio *et al.* (2023), Faure-Beaulieu *et al.* (2023)), with compressibility (Currao (2024), Kickliter *et al.* (2025)). The continuity and RANS equations in compressible index summation notation form are shown in equations 3.1-3.5, where $\rho$ is density, $t$ is time, $x_i$ are Cartesian coordinates for the purpose of writing these equations, $u_i$ is the velocity vector, $\tau_{eff,ij}$ is the deviatoric total stress tensor comprised of the sum of a laminar contribution ($\tau_{lam,ij}$) and a turbulent one ($\tau_{turb,ij}$), $P$ is the pressure, $C_p$ is the constant pressure heat capacity. Also, $T$ is the static temperature, $\gamma$ is the thermal conductivity, $\mu_{lam,m}$ is the molecular viscosity (shown here as Newtonian for simplicity but is later addressed as non-Newtonian in table2), $F_{s,i}$ is the surface tension force due to interfacial curvature, $i\,j\,k$ are summation indices, $\delta_{ij}$ is the Kronecker delta, $\Pi$ is fuel volume fraction, $\mathcal{k}$ is turbulence kinetic energy, and all subscripts $m$ designate mixture properties which have been volume weighted by phase volume fraction. Using the traditional VOF approach with PLIC for SGS liquid-gas interfacial calculations, the solution to a single velocity and temperature field are sought.

$$\frac{\partial \rho_m}{\partial t} + \frac{\partial}{\partial x_i}(\Pi \rho_m u_i) = 0 \tag{3.1}$$

$$\frac{\partial}{\partial t}(\rho_m u_i) + \frac{\partial}{\partial x_i}\left(\rho_m u_i u_j\right) = \frac{\partial \tau_{eff,ij}}{\partial x_i} - \frac{\partial P}{\partial x_i} + F_{s,i} \tag{3.2}$$

$$\rho_m C_{p,m} \frac{\partial T}{\partial t} + u_i \frac{\partial T}{\partial x_i} = \gamma_m \frac{\partial^2 T}{\partial x_i^2} \tag{3.3}$$

$$\tau_{lam,i} = \mu_{lam,m} \left( \frac{\partial u_i}{\partial x} + \frac{\partial u_j}{\partial x_i} \right) - \frac{2}{3}\mu_{lam,m} \frac{\partial u_k}{\partial x_k} \delta_i \tag{3.4}$$

$$\tau_{turb,i} = \mu_{turb,m} \left( \frac{\partial u_i}{\partial x} + \frac{\partial u_j}{\partial x_i} \right) - \frac{2}{3}\rho k \delta_i \tag{3.5}$$

$\mu_{turb,m}$ and $k$ are computed by the well-documented SST turbulence model. The "scalable" near-wall treatment adapts to the local near-wall cell size based on local $y^+$ values to ensure that the first cell center is appropriately handled for wall shear stress; $F_{s,i}$ and $\Pi$ are discretized and calculated based on the well-documented PLIC method. Contributions by viscous heating, pressure-work, and kinetic energy have been neglected from the energy equation.

### 3.2. *Mesh*

Two full 360° nozzle meshes were considered herein as is provided by figure 3, where flow is generally from left to right. The purely black areas in the figure represent the solid makeup of the atomizer body. The first mesh was based on "R1" from Wilson & Strasser (2022*b*) at 8.76 million cells. This mesh was designed to resolved enough physics to develop the correct WAVE behavior but not fine enough to be mesh-independent. Droplets were reasonably captured but were too large in "R1". A cross-section along the atomizer center plane is provided on the LH of figure 3.

The second mesh was based on the mesh-independent "R3", also from Wilson & Strasser (2022*b*), at 152 million cells. Its center place is shown on the RHS of figure 3. There are two significant differences between the two meshes in this figure. The first is that the RHS mesh is refined twice more (by halving the element length scale in all three dimensions each time). The second is that the refinement is isolated to the area of physics interest where the fuel and gas meet to develop those instabilities necessary for direction-switching tangential motion. In both mesh cases, the area of interest contains near-unity aspect ratio hexahedral cells with no cut-cell boundaries. Automatic and localized mesh refinement was ruled out due to potential inaccuracies described in Wilson *et al.* (2023*a*). In addition to the R1 360° mesh shown on the LHS of figure 3, some tests herein will be carried out to assess the impact of azimuthal angle on the results. Both an R1-90° and an R1-180° meshes were considered. Of course, those mesh element counts would be the commensurate fractions of the full 360° version.

### 3.3. *Boundary Conditions*

Atomizer dimensions, properties (including the shear- and temperature-dependent Herschel–Bulkley viscosity parameters for the fuel), and boundary conditions are given in Wilson & Strasser (2022*b*) but are summarized for the reader in upcoming tables 1 and 2. Implicit by the tables showing mass flow rates, the inlets were constant mass flow boundaries. Outlets were set to constant atmospheric pressure, and all walls were no-slip. For partial azimuthal simulations, the azimuthal bounding faces were periodic. Initial conditions, as described elsewhere, were inherited from prior quasi-steady state (QSS) models presented in Wilson & Strasser (2022*b*). Developing scaling relationships for our newly proposed behavior using key dimensionless numbers (see table 5 in the Appendix) would be interesting. An example might be to vary mass flows of the fuel and steam to assess swirling strength. However, the simulations described herein already required enormous resource allocation. This renders other boundary condition evaluations inaccessible at the present.

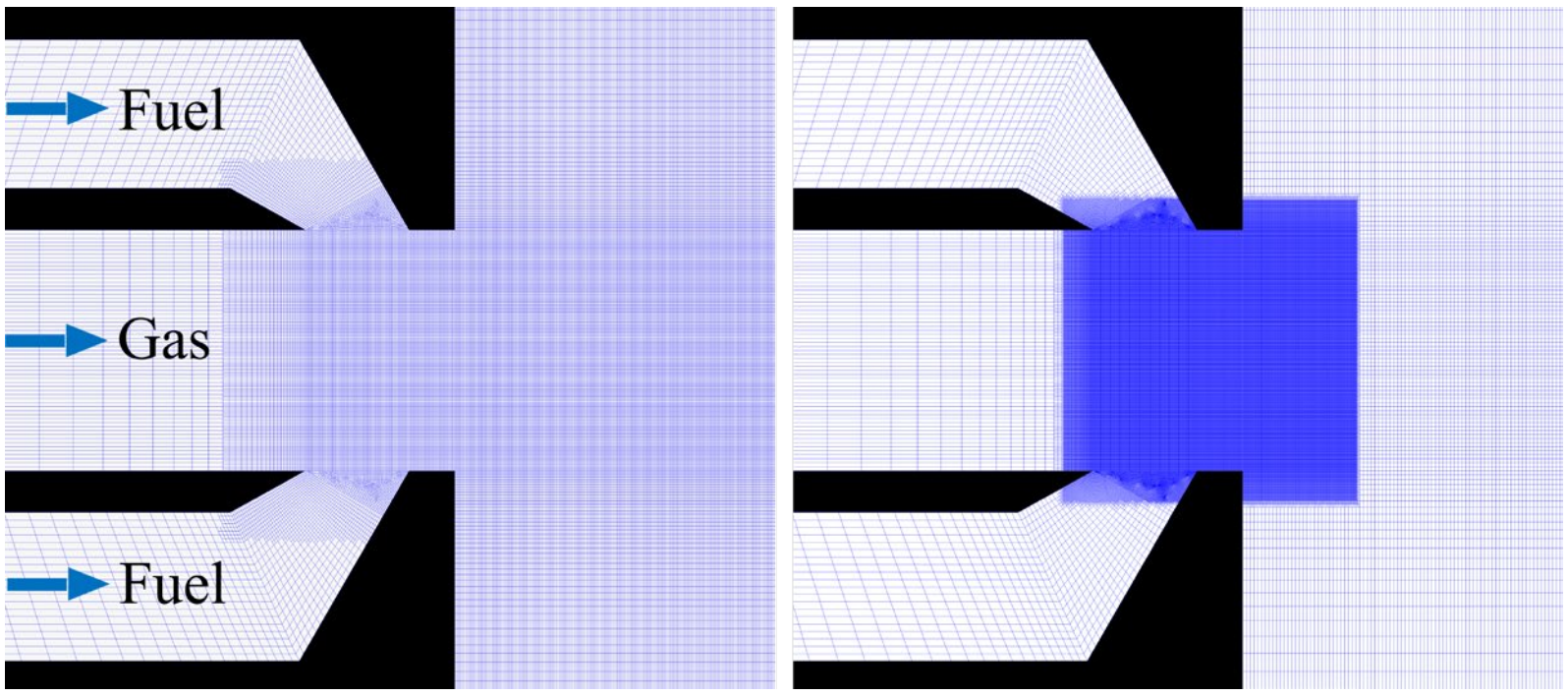

Figure 3: Center planes of a relatively coarse 360° mesh ("R1", 8.76 million hexahedral elements, left) and our fine 360° mesh ("R3", 152 million hexahedral elements, right) used for the majority of work herein. Flow is generally from left to right. The model's inlets (left), outlet (right), and wall boundaries are beyond what is shown in these images. The atomizer is geometrically axisymmetric, but the flow far from it. Not shown are the R1-90° and R1-180° meshes also considered.

| Steam (Core) Property | Symbol | Value (SI) |
|---|---|---|
| Feed Flow Rate | $\dot{m}_{f,1}$ | 0.0210 |
| Molecular Viscosity | $\mu_1$ | $1.79 \times 10^{-5}$ |
| Density | $\rho_1$ | Ideal Gas |
| Heat Capacity | $Cp_1$ | 1890 |
| Thermal Conductivity | $k_1$ | 0.0250 |
| Feed Temperature | $T_{f,1}$ | 393 |
| Feed Sound Speed | $a_f$ | 488 |
| Heat Capacity Ratio | $\Gamma$ | 1.31 |

Table 1: Properties of steam, the core fluid, or fluid 1 based on data from Wilson & Strasser (2022*b*), Wilson & Strasser (2022*a*), Wilson *et al.* (2023*a*), and Wilson *et al.* (2023*b*). We ignore compressibility effects in the annular fuel, so the compressible properties for the core fluid are not assigned subscripts.

### 3.4. *Numerics and Validation*

Equations 3.1-3.5 were solved using double-precision Fluent 2020R1 with the Semi-IMplicit Pressure Linked Equations (SIMPLE) for segregated pressure-velocity coupling. PREssure STaggering Option (PRESTO!) for pressure discretization and second order upwind-biased discretization for the momentum equation and were employed. Time step was varied to maintain a globally maximum Courant number of $< 0.9$ throughout the entirety of the simulations. The time step size was $O(10^{-8}\ s$, which equates to one-hundred-thousandth of a wave/pulsation cycle. In addition to the aforementioned support from other researchers treating their equations likewise in Fluent, our past validation work including both mesh- and numerics-independence using these general approaches can be found in Strasser (2011), Strasser & Battaglia (2016), Strasser & Jiang (2024), and Strasser & Morgan (2025). Moreover, the globally pulsing character of similar injector designs are widely known (Sahoo & Gadgil (2022)).

| Fuel (Annulus) Property | Symbol | Value (SI) |
| --- | --- | --- |
| Feed Flow Rate | $\dot{m}_{f2}$ | 0.790 |
| Molecular Viscosity | $\mu_2$ | Herschel–Bulkley (Non-Newtonian) |
| Density | $\rho_2$ | 1030 |
| Heat Capacity | $Cp_2$ | 4180 |
| Thermal Conductivity | $k_2$ | 0.640 |
| Surface Tension | $\sigma$ | 0.07 |
| Feed Temperature | $T_{f,2}$ | 304 |

Table 2: Properties of fuel, the annular fluid, or fluid 2 based on data from Wilson & Strasser (2022*b*), Wilson & Strasser (2022*a*), Wilson *et al.* (2023*a*), and Wilson *et al.* (2023*b*). Herschel–Bulkley parameters can also be found in these.

| Time-Averaged (TA) Result or Dimension | Symbol | Value (SI) |
| --- | --- | --- |
| Core Gas Radius | $R_1$ | 0.00671 |
| Area-Averaged Core Gas Velocity | $W_1$ | 110 |
| Core Gas Centerline Velocity | $W_c$ | 341 |
| Core Gas Feed Velocity | $W_f$ | 79.3 |
| Orifice Radius | $R_2$ | 0.00790 |
| Steam Feed Diameter | $D_f$ | 0.0158 |
| Annular Fuel Thickness | $R_A$ | 0.00119 |
| Area-Averaged Annular Fuel Velocity | $W_2$ | 14.1 |
| Fuel Wavelength | $\lambda$ | 0.0107 |
| Fuel Wave Height | $h_W$ | 0.004 |
| Fuel Wave Time | $t_W$ | 0.001 |
| Fuel Wave Speed | $c_W$ | 11.0 |

Table 3: Relevant dimensions and time-averaged (TA) results to be used in dimensionless numbers in table 5. All are based on data from Wilson & Strasser (2022*b*), Wilson & Strasser (2022*a*), Wilson *et al.* (2023*a*), and Wilson *et al.* (2023*b*). $W_{c,1}$ is calculated based on TA CFD data at the orifice centerline (CL) and not based on a simple Hagen-Poiseuille flow as done by Kerchman (1995). If the latter method were used, the computed CL velocity is approximately 4 orders of magnitude too high. Also, AA fuel velocity is based on TA fuel annular thickness.

There are no experimental data with which to compare these numerical results; however, we have the aforementioned works which confirm some level of accuracy in our underlying numerical constituents. More specifically, Strasser (2011) confirms that the current method captures self-generating and self-exciting pulsing annular wave patterns and acoustics correctly for transonic multiphase flow. Strasser & Battaglia (2016) show that our droplet size analyses under shock-droplet shattering conditions matches known trends and that non-Newtonian droplet production is precisely captured. These are also summarized in Wilson & Strasser (2022*b*). Additionally, the viability of this approach was affirmed in Strasser & Jiang (2024). Finally, the underlying numerical scheme was more closely scrutinized in Strasser & Morgan (2025). It is acknowledged that no numerical tool is without flaw and bias. It is also known that numerical tools can induce instability (Prichard *et al.* (2026), Strasser &

Morgan (2025)). Therefore, it should be expected that our method is imperfect. On the other hand, based on the aforementioned prior validation efforts, it is unlikely that the observed SASS behavior is exclusively numerically sourced. This is especially true given that our tool has accurately captured two major physics embodiments involved here. The first is the self-generative (by known instabilities) wave motion at the correct frequency. If our numerical tool can correctly capture waves in the streamwise direction, why would it not be able to do the same for waves which are generated orthogonal to those; the underlying mechanisms and numerical burden are precisely the same. Secondly, we have captured experimentally verified droplets before. This is a critical element, even though we are not studying droplets in this paper, because the same VOF routine which explicitly forms droplets without excessive numerical smearing or dispersion, is also, by definition, able to capture a wave with much less local curvature than a micron-sized droplet.

More attention is now given to the SST turbulence model. It, along with any other linear eddy-viscosity model is certainly insufficient to capture 3D turbulence. In fact, all RANS models are incapable of capturing true turbulence; however, this does not mean that the SST cannot accurately approximate transient 3D flows and instabilities. The aforementioned validation work, for example, involved the SST in every case. As it is well-known, KHI and RTI are inviscid phenomena, meaning that turbulence could be irrelevant to their behavior. The reader will also be aware of the potential improvement in using large eddy simulations (LES) or hybrid RANS-LES. In fact, the authors have demonstrated validation in each of these, respectively, in Strasser (2022) and Strasser *et al.* (2024). At the same time, Pope (2001) warns that a poorly designed LES is worse than a simple mixing length model. We cannot, therefore, be sure that LES is necessarily better. Additionally, the authors have evaluated SST relative to Implicit LES in a yet-to-be-published effort and found the exact same frequency and pressure gradient pulsation standard deviations between the turbulence modeling frameworks, each to 2 significant digits. Wilson & Strasser (2022*b*) and Wilson & Strasser (2022*a*)discuss the facts that 1. an estimate for the integral length scale of turbulence is an order of magnitude lower than our droplet length scale and 2. turbulence is not the dominant player in wave formation. In summary, until this point, we have not found justification to warrant the substantially increased meshing demands for LES in these atomization systems. In fact, we demonstrate in this paper alone that the SASS response is, to a large degree, independent of a wide range of mesh resolutions.

Regardless, however, we recognize the uncertainty in our results and that an experimental test rig should be configured for further validation, but that is relegated to a future study. Since the effects detailed herein are 3-D in nature and only exist in non-planar geometries, the experimental apparatus would need to involve high-speed photography embedded within a curved channel. The camera frame rate should be at least 20,000 Hz to capture 20 frames per cycle in order to visually confirm the SASS mechanism. Other obvious validation points include pressures using transmitters on the feeds and interfacial velocity order-of-magnitude estimates from camera footage. Pressure transmitters would also be helpful for inlet steam pressure fluctuations.

### 3.5. *Computing Resources*

The 360° R3 simulation was executed on 15 compute nodes, each with 120 AMD Epyc 7003X-series (also known as "Milan X") cores for about 6 weeks, which translates to approximately $2 \times 10^6$ core-hours. The progression rate of the solver was, therefore, about 4 wave passes per million core-hours. These core-hours exclude months of running the 360° R1 mesh variants on smaller computer core counts. Data used to initialize the 360° R3 simulation was extracted from a QSS version of its 360° R1 counterpart simulation presented in Wilson & Strasser (2022*b*). The total flow time duration of the simulations, along with the flow time necessary for reaching QSS and then garnering sound statistical measures, will be a topic of discussion later in the paper.

## 4. Results

### 4.1. *Radial Modal Deformation*

Before the 3-D CFD results are discussed, we shall consider underlying physics on a less rigorous scale to set up a precursor for the next section on surface deformation. To do so, various governing equations in simplified form will be employed in thought experiments to examine *directional* patterns. No instability analyses are warranted, and our wave fluctuations are far too large to be encompassed by any linear stability theory. The continuity equation for an incompressible, steady, and laminar fuel-only differential shell in cylindrical coordinates is equation 4.1.Although our assisting gas flow is compressible in this design, its compressibility is irrelevant to our newly discovered mechanism.

$$\frac{1}{r}\frac{\partial}{\partial r}\left(r u_{2,r}\right) + \frac{1}{r}\frac{\partial u_{2,\theta}}{\partial \theta} + \frac{\partial u_{2,x}}{\partial x} = 0 \tag{4.1}$$

The velocity components and coordinate system are shown in figures 4 and 5, where $u_{2,r}$, $u_{2,x}$, and $u_{2,\theta}$ (not pictured) are the radial, axial, and azimuthal fuel velocity components, respectively. Figure 4 illustrates an instantaneous snapshot of the fuel volume fraction (yellow) on a nozzle center plane from 360° mesh "R3". A time-lapse animation of this figure from the full CFD simulation can be found here. (The yellow fuzziness on the RH side of the image is due to the sudden mesh size increase at that location and is beyond anything we care about in this paper.) After the fluids meet, the less dense and less viscous gas becomes the core material (fluid 1), and the denser and more viscous fuel becomes the annular material (fluid 2) to produce CAF. The close-up on the right includes not only the velocity components but also the wave identification; wave $n$ is the current wave passing through the orifice, and wave $n_1$ is the last wave which passed through the orifice. As discussed extensively in the prior cited papers, wave $n$ roles up due two causes. First, fuel deceleration (relative to wave $n$) in wave $n_ 1$ is caused by fuel drag on the solid orifice ledge, which is what we have called the beach. The second cause is vorticity induced by adjacent steam flow.

The idea with our simplified analysis is that a new fuel wave has "piled up", more so than "rolled up", on a subsequent wave. A snapshot in time is analyzed, such that the underlying flow is all steady, incompressible, and inviscid. In other words, the new wave is simply radially adding fuel material, all at the same axial velocity.

The outcome of applying the chain rule and then rearranging 4.1 is equation 4.2:

$$-\frac{\partial u_{2,x}}{\partial x} = \frac{\partial u_{2,r}}{\partial r} + \frac{u_{2,r}}{r} + \frac{1}{r}\frac{\partial u_{2,\theta}}{\partial \theta} \tag{4.2}$$

Since $\partial u_{2,x}/\partial x < 0$ via streamwise deceleration, the three terms on the right hand side ($RHS$) must sum to $> 0$. It will now be temporarily assumed that there is not *yet* any flow in $\theta$. The shell's outer impenetrable wall ensures $u_{2,r} < 0$, leaving the fuel flow pointing radially inward, and $r > 0$; therefore, $u_{2,r}/r < 0$. The only way the $RHS > 0$ is if $\partial u_{2,r}/\partial r > 0$, indicating radially inward spatial acceleration. Additionally $\partial u_{2,r}/\partial r > abs\left(u_{2,r}/r\right)$.

Now, we temporarily simplify the Navier Stokes equations to neglect transient effects (buoyant forces are already dismissed due to our high Froude number) and viscous forces. This produces the $r$-component of linear momentum (Euler equation) for a steady incompressible fuel:

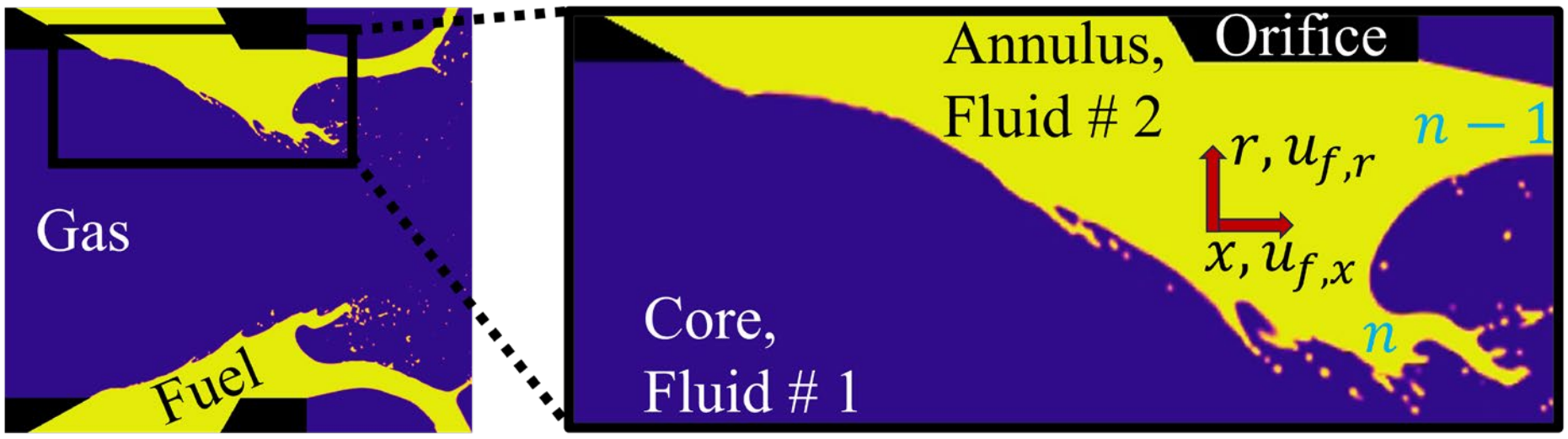

Figure 4: Nozzle cross-section from 360° mesh "R3" showing a close-up of the instantaneous gas phase (purple), fuel (yellow) wave identification, coordinate system, and velocity components of immediate interest. The fuel becomes the much more viscous and much more dense annular material (fluid 2), while the gas becomes the core material (fluid 1). Again, bulk flow is from left to right. Wave $n$ rolls up or piles up on wave $n - 1$. A time-lapse animation based on volume fraction from this view can be found here.

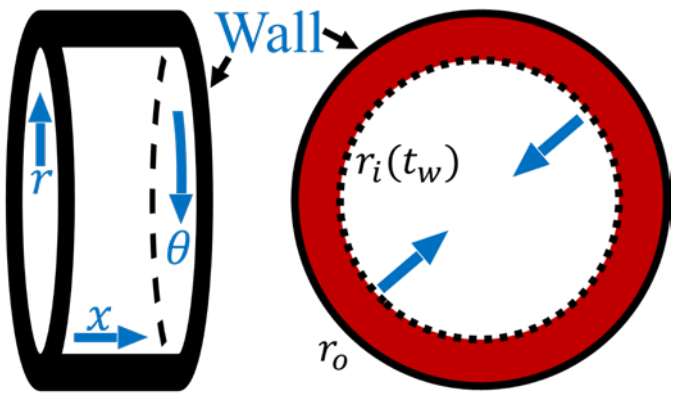

Figure 5: Two views of the same simplified conceptual differential cylindrical fuel shell immersed in the fuel wave from figure 1; the left image is an oblique view, where bulk flow is from left to right along the nozzle axis in $x$ at $r = 0$; the right image is viewing oncoming bulk flow (parallel with the nozzle axis). The red region on the right is the fuel annulus bound by the pipe wall on its outside but with an inwardly moving radial interface at $r_i$ due to wave pile-up.

$$\frac{\partial P}{\partial r} = -\rho_2 \left( u_{2,r} \frac{\partial u_{2,r}}{\partial r} + \frac{u_{2,\theta}}{r} \frac{\partial u_{2,r}}{\partial \theta} - \frac{u_{2,\theta}^2}{r} \right) \tag{4.3}$$

If, again, it is temporarily assumed that there is not *yet* any flow in $\theta$ for the purpose of making a single point, the result is equation 4.4.

$$\frac{\partial P}{\partial r} = -\rho_2 u_{2,r} \frac{\partial u_{2,r}}{\partial r} \tag{4.4}$$

It has already been established that $\partial u_{2,r}/\partial r > 0$ and $u_{2,r}/r < 0$; therefore, $\partial P/\partial r$ must be $> 0$. This implies pressure falls with decreasing radius. An important clarification is that the sign of the radial pressure gradient is irrelevant, because it is based on our $r$-axis convention. The fact that pressure falls along the flow direction makes $\partial P/\partial r$ a *favorable* pressure gradient. While this confirms our intuition from Bernoulli, it points out an important feature: there is a surviving radial pressure gradient in the fuel that is defined by combining equations 4.2 and 4.4 to form:

$$\frac{\partial P}{\partial r} = \rho_2 u_{2,r} \left( \frac{\partial u_{2,x}}{\partial x} + \frac{u_{2,r}}{r} \right) \tag{4.5}$$

In other words, the merging of fuel waves decelerates the fuel in the axial direction. This produces a radial fuel influx which drives its own radial pressure gradient. The radial pressure gradient then scales with $\sim 1/r$, holding the axial velocity gradient constant and neglecting flow in $\theta$. Augmenting this process is the input energy source of steam and fuel enthalpy. In addition, kinetic energy is supplied by the fuel and steam feeds and can be generally thought of as the driving pressure $\partial P/\partial x$.

Instabilities later to be discussed then trigger infinitesimal $\theta$ motion. This returns terms 2 and 3 originally on the RHS of equation 4.3 and now must be included in equation 4.5 as shown below in equation 4.6.

$$\frac{\partial P}{\partial r} = \rho_2 u_{2,r}\left(\frac{\partial u_{2,x}}{\partial x} + \frac{u_{2,r}}{r} - \frac{u_{2,\theta}}{r}\frac{\partial u_{2,r}}{\partial \theta} + \frac{u_{2,\theta}^2}{r}\right) \tag{4.6}$$

Considering those new terms on the RHS of equation 4.6, one notices a mixed effect of $\theta$-related flow on the radial pressure gradient. First, the radial fuel pressure gradient is reduced by the product of $\theta$-velocity and the gradient of $r$-velocity in $\theta$; therefore, as inward-facing fuel velocity develops azimuthal variation, the radial pressure gradient is suppressed. On the other hand, the $\theta$-aimed velocity, in general, non-linearly exacerbates the fuel's radial pressure gradient.

Let us now introduce the time-varying character of the fuel annular thickness via $r_i(t)$ in figure 5 via equation 4.7. In this case, $t$ operates at wave time scales (see table 3) which are large compared to the integral time scale of the largest eddies modeled by RANS ($\propto k/\varepsilon$, where $\varepsilon$ = dissipation rate). Therefore, the traditional "separation of scales" is maintained between RANS turbulence and wave motion.

$$r_i(t) = \sqrt{r_0^2 - \frac{A(t)}{\pi}} \tag{4.7}$$

Variable mass flow through the fuel control volume by a wave pulse is given by:

$$\dot{m}_2(t) = \int \rho_2 \mathbf{u}_{2,x} \cdot d\mathbf{A}(t) \tag{4.8}$$

Taking $\mathbf{u}_{2,x}$ to be normal to $\mathbf{A}(t)$, independent of $\mathbf{A}(t)$, and approximately constant with time, equations 4.8 and 4.7 become equation 4.9. Stated differently, the annular fuel wave moves at a relatively constant axial velocity, while the *thickness* of the annulus is growing with time (wave "pile-up"). The pile-up advances the annular interface inward into the gas flow.

$$r_i(t) = \sqrt{r_0^2 - \frac{\dot{m}_2(t)}{\pi \rho_2 u_{2,x}}} \tag{4.9}$$

We next postulate that the radially and inwardly advancing fuel annulus front will disrupt the favorable $\partial P/\partial r$ established by the existence of $u_{2,r}$ in 4.6. This happens for at least two reasons. First, if considering the presence of the steam core, its pressure is likely higher (in a sinusoidal fashion) than that of the fuel front. Second the fuel wave piling up continues to crowd itself while moving radially inward. It can be concluded, then, that $\partial P$ will rise as $\partial r$ falls, making the pressure gradient *adverse* and working against that already established. The normal pressure development is expected to scale with the time derivative of $r_i(t)$ given by equation 4.10 below. This occurs as the fuel resists being accelerated out of the way of more incoming fuel flux:

$$\frac{dr_i(t)}{dt} = -\frac{1}{2\pi\rho_2 u_{2,x}} \sqrt{\frac{\pi\rho_2 u_{2,x}}{\dot{m}_2(t)}} \frac{d\dot{m}_2(t)}{dt} \tag{4.10}$$

Then, when we take $\dot{m}_2(t)$ to be some sinusoid, equation 4.10 takes on the form below, where $c$ is some constant that is unimportant to this line of thought. Therefore,

$$\frac{\partial P}{\partial r} \propto \frac{dr_i(t)}{dt} \approx -c\sqrt{\frac{1}{sin(t)}}\,cos(t) \tag{4.11}$$

The radial adverse pressure gradient is the opposite in sign of the prior discussed Bernoulli-like favorable pressure gradient. If there were a contraction in the orifice boundaries ($\partial r_0/\partial x < 0$) or in time ($\partial r_0/\partial t < 0$), the fuel radial deformation would be exaggerated.

The $r$-momentum equations cannot be truly assessed in isolation, because, by the continuity (equation 4.2), any $\theta$-induced velocity will mitigate radial or axial fuel velocity. However, for the purposes of approximation general effects, we are neglecting this. What cannot be neglected, though, is that $\theta$-based flow engages the $\theta$-component of linear momentum for a steady, incompressible fuel in equation 4.12:

$$\frac{\partial P}{\partial \theta} = -r\rho_2\left(u_{2,r}\frac{\partial u_{2,\theta}}{\partial r} + \frac{u_{2,\theta}}{r}\frac{\partial u_{2,\theta}}{\partial \theta} + \frac{u_{2,\theta}u_{2,r}}{r}\right) \tag{4.12}$$

It is evident that the azimuthal pressure gradient is determined by $\theta$-velocity and its spatial gradients (particularly self-acceleration, the second RHS term), along with the same $r$-velocity that initially produced the radial pressure gradient. Since pressure at a point inside the differential shell (figure 5) must be balanced in all directions, any introduction of $\partial P/\partial\theta$ will necessarily be felt by $\partial P/\partial r$ (and $\partial P/\partial x$, not shown). In short, instabilities generating $\theta$-flow will, in turn, perturb $\partial P/\partial r$, which will create spatiotemporal variation in $u_{2,r}$ and will continue the instability cycle.

To summarize this subsection, the merging of waves produces a fluctuating radial fuel pressure gradient. Bernoulli implies that this gradient is favorable; while an advancing fuel front suggests it would be adverse. These effects are adjusted by any perturbations in $\theta$. We openly recognize that, in seeking conceptual analysis to establish the precursor pressure gradient discussion for our upcoming surface deformation section, the kinematics of the system are too complicated to analytically resolve the aforementioned expressions. Additionally, we have neglected the following: interplay between $u_{2,r}$ and $u_{2,x}$ in the advancing fuel front, in-plane Newtonian and non-Newtonian viscous stresses, strong temporal disturbances in all directions of motion, and the implications of destabilizing character of surface curvature on turbulence. Consequently, any future analysis will be based on CFD results of the solution to equations 3.1-3.5.

### 4.2. *Slender Lobe Production*

All upcoming discussion will be supported by data and figures later in the article. With the introduction of a conceptional analysis in the prior section, we now shift focus to the mechanisms which create spontaneous axis-switching swirling (SASS) behavior. It begins with slender lobes resulting from the aforementioned radially and inwardly advancing fuel front. As the fuel piles up on itself (think of the radial deformation of a metal cylinder), the inner enclosed fuel surface diameter continues to fall. This raises pressure locally, and that pressure must be relieved somehow. Relief is made possible by a surface imperfection allowing the annulus to “buckle” in a sense. slender lobes are the result of the fuel cylinder buckling and is illustrated in figure 6 made up of three images. The upper LH picture shows an undisturbed fuel annulus running between the steam core and the nozzle interior. The lower LH picture shows newly emerged slender lobes, and the RHS shows a close-up of the slender lobes exiting the nozzle. None of these fuel annuli images include the primary axial waves developing from KHI. Those base streamwise waves are certainly there and active; they have only been excluded for simplification. The new slender lobes grow, i.e. lifting farther off of the surface moving downstream. They are also developing their own wave pattern orthogonal to those streamwise surface waves not shown. Depicted here is the reduction in slender lobe wavelength, and an increase in amplitude, towards the exit. Because the fuel is at a much higher density than the core steam, the slender lobes survive a finite distance before being stripped off of the fuel annulus by the high-speed gas core.

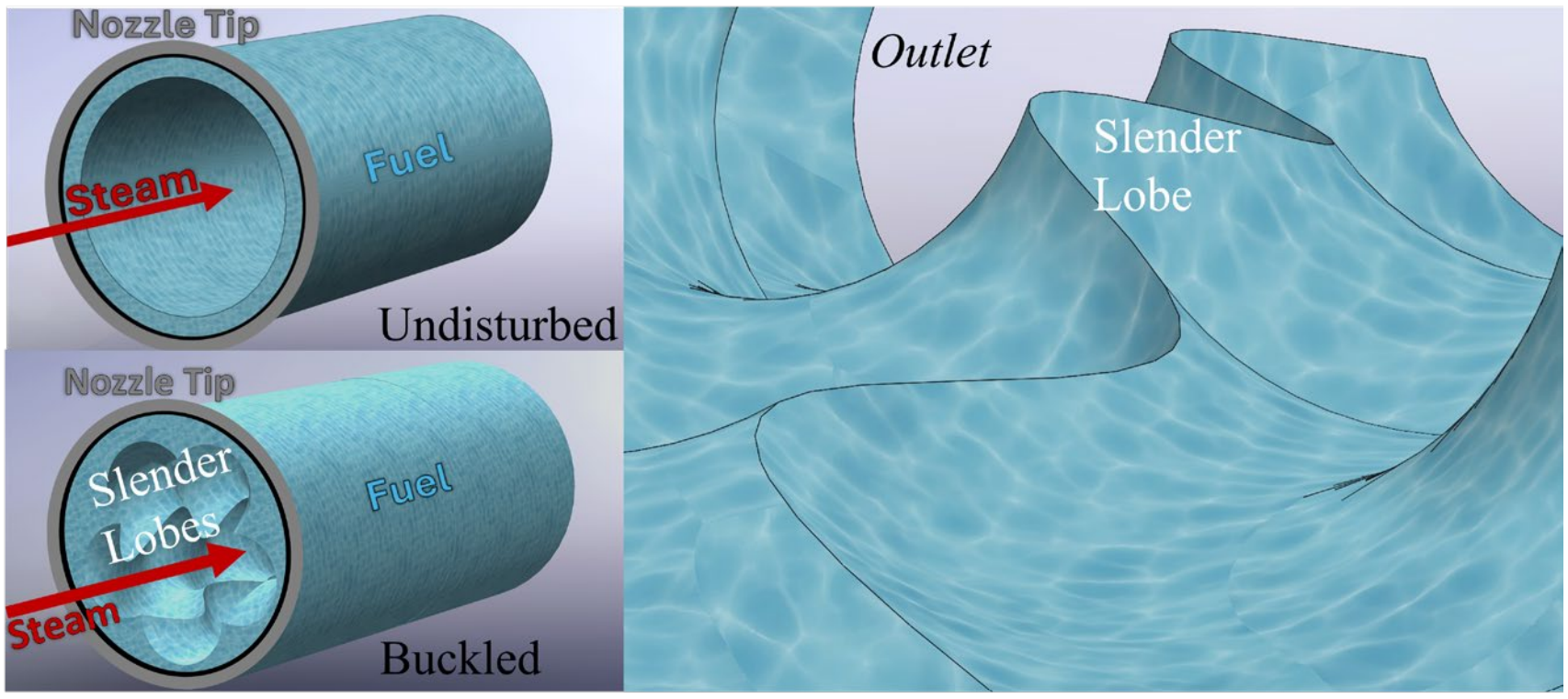

Figure 6: Conceptualization of slender lobes forming on the fuel annulus inner surface. The upper LH corner shows an undisturbed fuel annulus, the lower LH corner shows the annulus after deformation, and the RHS shows the inner ridge (towards the outlet) developing an increasing wave amplitude in a direction orthogonal to the original base flow. Flow is generally southwest to northeast.

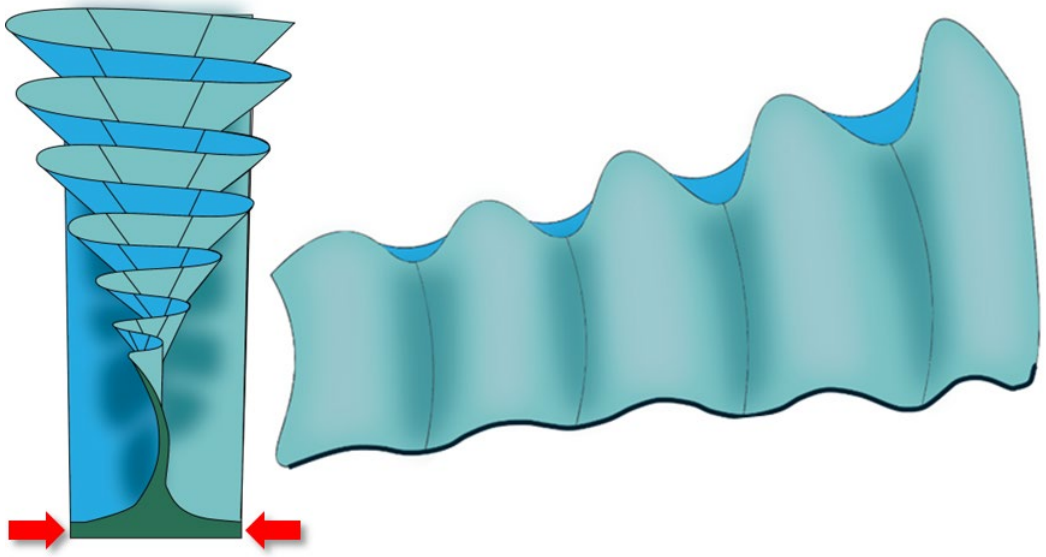

Figure 7: Additional views of the conceptualized and exaggerated fuel slender lobe growing due to KHI which are orthogonal to the base flow. The LH image is an overhead view of an ultra-thin and tall slender lobe, and the RH image is a side-view of the same slender lobe. Red arrows show the azimuthal influx of fuel feeding the slender lobe. The reader will recall that moving up in the LH image is the bulk flow streamwise direction.

Two additional views of a single slender lobe structure are shown in figure 7. The LH image considers an overhead view with flow from bottom to top, and the RH image is a side view with flow from left to right. In both cases, the slender lobe itself has been exaggerated to be quite tall and thin for the purpose of discussion. Red arrows are provided to show the azimuthal influx of fuel which provides influx at the base of the slender lobe. As with figure 6, the slender lobe is shown growing in height and wave amplitude with axial flow distance.

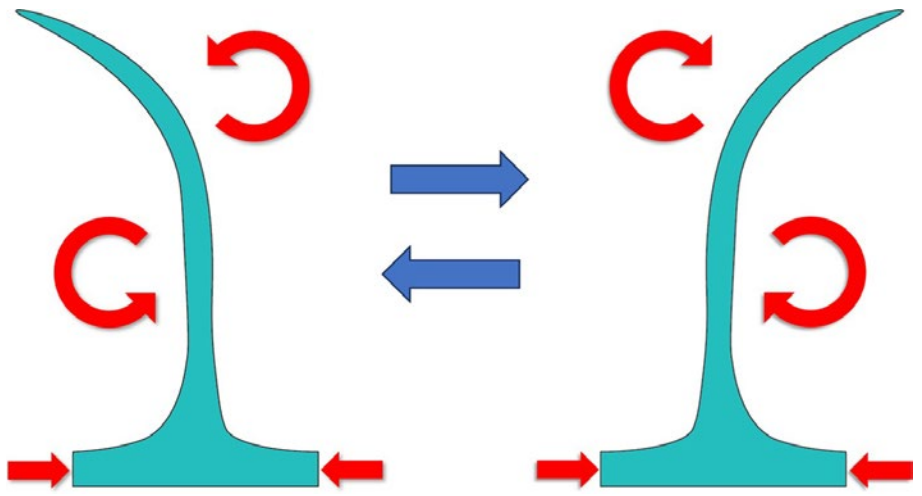

Figure 8: A single fuel slender lobe taken from a infinitesimal slice from 7 is being buffeted laterally in $\theta$ due to its wavy structure. The LH and RH images are the same slender lobe but just bending in opposing directions at different times, indicated by the two opposing horizontal blue arrows. Additional red curved arrows at the top of the slender lobe illustrate secondary steam flows which provide additional stimulation. Looking into the page is the bulk flow streamwise direction.

The cause of the slender lobe shape being deformed into newly emerging waves is conveyed with the assistance of figure 7. As the slender lobe rises from the surface, it effectively divides the parallel gas flow into left and right streams. Considering the LH side of figure 7, there is a high-speed gas on both its left and right after bifurcation. (By "left" and "right" in this case, we are referring to different positions in the annular azimuth, or $\theta$.) In all likelihood, those two gas streams are not the same speed, especially given the non-uniform pulsing turbulent gas core. The two differing speeds on either side of the slender lobe produce KHI which causes the slender lobe to "wind-up" as shown in both the overhead and side views.

The wavy $\theta$ motion in 7 is further examined in 8, and this is where RTI become important. This figure was generated by taking a very small streamwise slice from the wavy slender lobe shown in 7. The tip bends left and right (in $\theta$) as the fuel experiences flow in the streamwise direction. The two blue horizontal arrows indicate direction-switching motion. During different times, the slender lobe is bending in opposite directions. The cycle repeats, which is what is implied by the oppositely facing arrows. Red curved arrows on either side indicate secondary steam flows adjacent to the laterally alternating tip structure. As the significantly less dense steam buffets the dense fuel, RTI emerge. They will produce bending and flexure of the slender lobe tips in the streamwise direction, i.e. ripples on the tip flowing along with the bulk steam. At that point, those ripples will likely compete with even newer KHI as the parallel flowing steam interacts with the ripples.

### 4.3. *CFD Evidence of Slender Lobes*

This section will involve instantaneous CFD images and hyperlinked time-lapse animations from on cross-cutting sampling planes like that shown in figure 1. Integer plane identifications (ID) will be given, and their axial location can be computed by equation 4.13, where $x_0$ is the axial location where the fuel and steam first meet; this would be considered the beginning of the "wave pool".

$$\frac{x - x_0}{r_0} = 0.1053 * ID \quad 0.0846 \tag{4.13}$$

As an example, figure 2 is based on Plane # 15 ("P15") which has an axial location of $x - x_0 \; r_0 = 1.66$, or 1.66 nozzle radii from the pool beginning. Planes 7, 9, and 11 are located at the beginning of the beach, middle of the beach, and end of the beach, respectively. In other words, planes 7 through 11 incorporate the flow within the nozzle orifice region.

Slender lobe precursors begin early in the nozzle as shown in figure 9 from 360° mesh "R3" looking directly into the flow at P4, where the dark colored gas core is seen surrounded by the fuel annulus. Outside the fuel is the nozzle inner surface (not shown). At this point in space and time, the inward radial fuel deformation produces relatively calm azimuthal interfacial perturbations. In viewing the time-lapse animation of this figure found here, the perturbations radially oscillate inward and outward as the streamwise waves pass; they also change shape and amplitude. However, they exist at relatively low length scales and don't yet emerge as slender lobes.

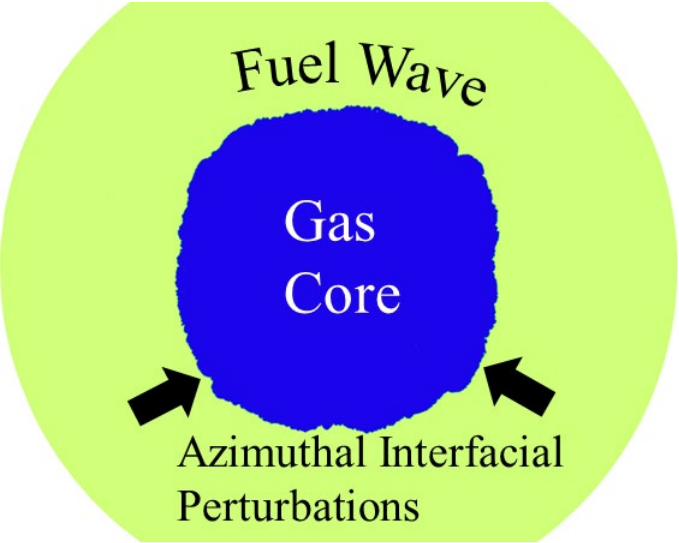

Figure 9: Instantaneous contours of fuel volume fraction on P4 from 360° mesh "R3", looking directly into the oncoming flow. A time-lapse animation with different colors but the same quantities and planar sampling location can be found here.

Moving downstream and including more metrics provides figure 10 also from 360° mesh "R3". There are 12 images total, all at the same (but arbitrary) snapshot in time. The LH column is volume fraction (dark is gas), while the next is strain rate (linear colormap from dark = 700 to light = $7 \times 10^5\, 1/s$). The third column is the Mach number (linear colormap from dark = 0 to light = 0.5), and the far RH side is tangential velocity magnitude (linear colormap from dark = $-20$ to light = $20 m/s$). The top row is based on axial location P9 (middle of the beach). The middle row is P11, and bottom row is P16. Farther upstream and sitting in the nozzle at P9, the fuel wave is surrounded by black indicating no computational domain, i.e. outside of the nozzle. The same can be said about P11 (end of the beach). At P16, however, the flow has just escaped the end of the nozzle orifice, and the fuel outer boundary becomes free to move. A time-lapse animation of each of these images in order of top-to-bottom (the volume fraction series first, i.e. "VF9", "VF11", and "VF16") and then left-to-right can be found linked directly in these shorthand designations: VF9, VF11, VF16, SR9, SR11, SR16, MN9, MN11, MN16, TV9, TV11, and TV16.

Beginning with the LH column of figure 10, and only considering the included still images for now, we see continued growth of those same perturbations initiated in figure 9. Obvious slender lobes have emerged between P4 and P9 and continue to radially protrude and thin by P11. On P16, the slender lobes have developed additional ligaments which are then subject to radially oriented KHI. The next column illustrates the strain rate which, as expected, is highest in the immediate vicinity of the fuel-steam interface. The very low strain rate regions in the fuel make a clear outline which matches the shape of the volume fraction interface. The innermost steam core remains at a relatively low shear due to lower velocity gradients. However, there is a comparably thick boundary layer representing the high-strain region in the steam flowing next to the fuel where turbulence generation occurs. In some locations, there are regions of low steam strain located between the fuel surface and regions of high steam strain. Those high steam strain areas are from separated shear layers off us upstream waves coming towards the reader. Column 3 entails what would be expected of a velocity profile via three characteristic features: 1. very low Mach number near the fuel surface, 2. relatively homogeneous steam core far from the fuel, and 3. a graduated boundary layer in between. Finally, in column 4, regions of positive and negative tangential velocity alternate in space around the periphery of the fuel wave. In some locations the direction is positive on one side of the slender lobe and negative on the other. The purpose of the somewhat low and arbitrary colormap upper limits ( |20 ) is to accentuate the spatial azimuthal *directional* changes (light next to dark); absolute values of tangential velocities, along with their temporal variability, will be visited in the next paper section.

Upon viewing the time-lapse animations referenced in figure 10, one finds similar features to what is included by comparing the three rows in the figure. The 2nd row is farther downstream than the 1st, and the 3rd is farther downstream than the 2nd. The SASS process is quasi-periodic, and the annular flow does not expand much between these planes; therefore the Eulerian and Lagrangian reference frames share comparable features. The ”VF9” animation reveals a cyclical pattern of radial expansion of the inner fuel interface and steam core with many slender lobes around the periphery pointing radially inward. Those slender lobes grow in number and are reduced in height as time passes and the fuel annulus thins. They also exhibit SASS behavior as the tips meander in alternative directions in the azimuthal dimension. The relative time spent in the full expanded (thinnest annular fuel with the most slender lobes) state is relatively larger than the time spent of the fuel interface expanding. Since plane 9 is located at the middle of the beach, the viewer will notice periodic occurrences of a double annular ring of fuel. This occurs when a fuel wave rolls over its predecessor, trapping the steam inside between the two fuel annuli and producing a steam annuli. The scene at VF11 (end of the beach) is similar to that of VF9. The primary difference is that the slender lobes and steam annulus are exaggerated. VF16 shows even more exaggerated behavior as the fuel just leaves the orifice. Its external boundaries are free to move. Two new elements arise. The first is that the fuel’s outermost boundary begins to deform, showing the early onset of new slender lobes. The most striking feature, however, is that the trapped steam annulus between sequential overturning waves begins to develop SASS at its outer periphery; those still point radially inward, however. The earlier animations were too constrained to enable this dual SASS behavior.

The animations tied to the second column (strain rate) show consistent behavior with the first column. As expected, the shear rates are always the highest at interfaces between the steam and fuel. That is where the high-speed steam is being decelerated by the viscous fuel. An important distinction for this column of animations is that there are large regions of speckled high strain in the wake of the droplets shed off the inner fuel wave surface at each planar location throughout the time-lapse. This behavior is most pronounced when the liquid annular is the thinnest. Evidently, that is the time in which a wave has just passed and fuel is being entrained off of the next wave’s valley. As with VF16, flow features are most obvious at SR16. Therein, the viewer can also see the droplets being shed from the interface between the outer liquid annulus and trapped steam. Column 3 (Mach number) shares similar features already discussed. Those, however, display the impact of steam flow cross-sectional area on steam Mach number. As the fuel interface expands, the flow area opens, and Mach number falls. The steam is rather homogeneous across its core. With animation MN16 where the fuel is outside the orifice, steam is being accelerated to appreciable Mach numbers by the outer fuel interface that has been accelerated by the inner steam. Finally, column 4 depicts azimuthal (tangential) velocity magnitudes at the various steam-fuel interfaces versus time. As with the 2nd column, the velocities are all dominant at the interface where slender lobes develop and weak elsewhere. As the fuel annulus thins and the steam core expands, those alternating patterns increase in frequency. The ratio of cross-sectional area occupied by highly active SASS increases moving along nozzle axis and in time as the steam core expands. In the latest stages of wave crashing and outer fuel interface energy in animation TV16, much of the visual landscape is occupied by energetic tangential oscillations. The visual effect results from a combination of two facts. First, more of the fuel is, indeed, being energies by SASS. Second, the chosen colormap boundaries have an effect. The low and high were chosen to visually accentuate the two different tangential directions. As the tangential *magnitude* increases, those boundaries would ideally be increased.

Additional time-lapse animations, but with images not shown herein, can be found for these three planes including tangential velocity vectors in two formats. The first will be called "FLV" for fixed-length vectors. For these three animations, the vectors are shown as arrows whose shaft length is fixed and with only arrow *colors* representing their magnitudes. Their colormap matches that of the aforementioned contours. The second will be referred to as "VLV" for variable-length vectors. These three animations involve vectors represented as arrows with *both* the color and the length reflective of tangential velocity magnitude (same color range as others). The downside to FLV is that stagnant fluid shows vectors, which crowds the scene and distracts from the primary motion. On the other hand, VLV sometimes has just a few very large arrows which obscure other primary flow. Representative time-lapse animations can be found linked directly to the following shorthand designations: FLV9, FLV11, FLV16, VLV9, VLV11, and VLV16.

In case the reader does not have access to the animations, they will be described here in the order provided above. Each of the three sets of videos (FLV9,11, and 16 along with VLV9,11, and 16) look similar to 4th column in 10. An obvious difference in FLV9 compared with TV9 is that the relative calmness of the flow outside of the SASS region (steam-fuel interface) is more clearly visible with the vector plots. Though it is not proof, it does reinforce the idea that the SASS mechanism is not purely numerically sourced. The SASS behavior follows the slender lobes and interfaces. If it were purely numerical in origin, it would be seen throughout the modeled domain as a result of too large of a time-step or related issue. The calm regions emerge and disappear with the passing of the interface, hinting at a lack of error propagation. Additionally, rotating structures (vorticity of these would be aligned with the viewing vector) emerge near the interface from the slender lobe tangential oscillation. One more unique aspect of the vector-based animations is the waves of coherent flow direction moving around in the steam core. There are patterns resembling stitching seams where the flow from different directions stagnate radially and move in a direction aligned with the viewing vector. This patterns moves with time and varies in the number and shape of the stagnation regions. Likewise, with the VLV9 animation relative to the other plane 9 animations, it is apparent that the flow outside of the steam core and steam-fuel interface is exceptionally calm, as expected, since this flow is mostly comprised of slower, more viscous liquid. The alternating SASS behavior at the slender lobe tips is also more pronounced here relative to the contour plot, along with the previously mentioned in-plane rotating structures at the bending slender lobe tips. The vector animations at planes 11 and 16, just like the contour animations, are mostly more active versions of their plane 9 counterparts.

In all of the images and time-lapse animations thus far referenced in this section, only a 2-D slice is visible, rendering what is occurring immediately upstream or downstream of these images invisible. To understand the more important 3-D character of the slender lobes, we consider figure 11, wherein the reader is viewing an iso-surface of fuel volume fraction at four progressing different arbitrary times, all looking from the outlet back into the flow. The time-lapse animation can be found here. Important to recognize is the fact that the iso-surface axial dimension (towards the reader) has been limited to include only the region containing the finest mesh elements shown in figure 3 and does not include any fuel which exists at or beyond the coarsened mesh interface. Annotated in the image is the remnant of previous wave $n_{-}1$ near the outer radial periphery, the dominant wave cresting in these images $n$, and a new wave $n_{+}1$ that is on its way to the forefront. Though not as orderly in CFD, the previously discussed slender lobe behaviors of slender lobe emergence and buffeting are demonstrated by wave $n$ at all four timestamps shown. Due to the fact that the 4 still images shown here are in time sequence, the animation provides the visual connectivity to smoothly transition between the still images. The viewer would see wave $n_{-}1$ move out of view, at which point in time wave $n$ fully consumes the visible fuel annulus. When this occurs, there is only one set of SASS slender lobes. Before this occurs, when the steam annulus exists between the waves, two sets of SASS slender lobes are evident in concentric radial rings. One is on the interface between the steam and wave $n_{-}1$, and the other is at the interface between steam and wave $n$. Both sets of slender lobes are protruding radially inward. Consistent with the prior animations is the radially expanding fuel interface and annular thinning as waves pass through and make room for more behind them.

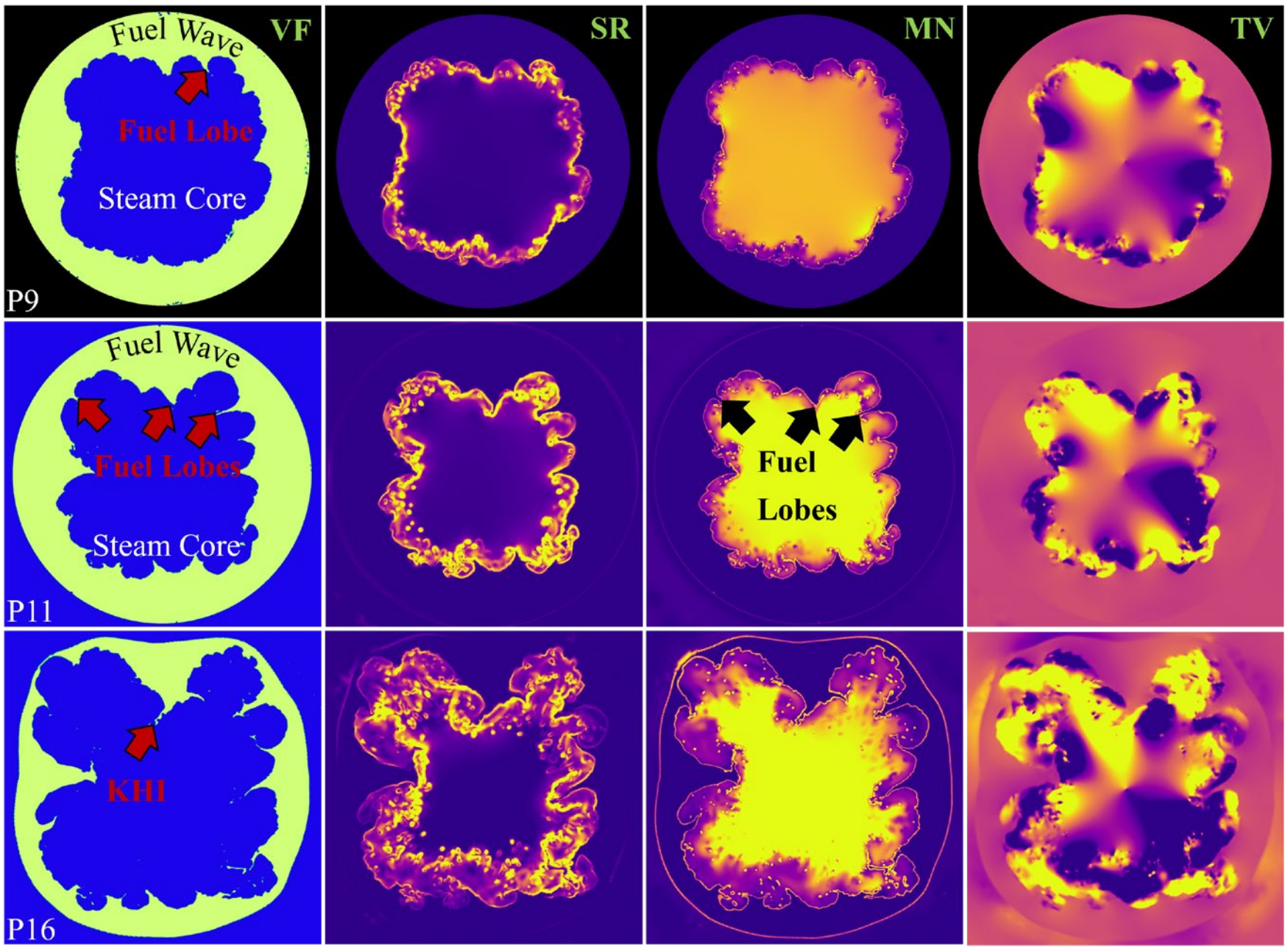

Figure 10: Instantaneous contours of fuel volume fraction ("VF", column 1), strain rate ("SR", column 2), Mach number ("MN", column 3), and tangential velocity component magnitude ("TV", column 4) on P9 (top row), P11 (middle row), and P16 (bottom row), all looking directly into the oncoming flow and from 360° mesh "R3". A time-lapse animation of each of these images can be found linked directly in these shorthand designations: VF9, VF11, VF16, SR9, SR11, SR16, MN9, MN11, MN16, TV9, TV11, and TV16.

Further facilitation of demonstrating the proposed slender lobe behavior in CFD is given in figure 12. Unlike the previous image, this fuel iso-surface view is made slightly oblique and is based on six progressive arbitrary times. Flow is generally from northwest to southeast. White arrows annotate the ripples which develop and then cause further azimuthal deformation. At time $t_1$, two slender lobes are marked. While irregular deformations are prevalent in both slender lobes, at $t_2$ a prominent wave is seen in the LH slender lobe. By time $t_5$ the wave has dissolved, and the upper ridges of both slender lobes are being buffeted to the left. The sequence ends with both of those slender lobes nearly out of the frame and a larger one has peeled off the side of the original RH slender lobe. Of course there is plenty of other activity in the images, include the traditional mushroom shapes resulting from RTI. Additionally, fuel droplets are seen throughout, as steam shear strips fuel off of the deformed interface. A broader view time-lapse animation is found here, wherein 25% of its azimuthal content has been removed to assist in viewing 3-D slender lobe behavior. Since figure 12 provides time-sequenced still images, its corresponding animation aims to include dozens of intermediate images between the six shown. slender lobes smoothly develop, following a reasonably straight path moving from northwest to southeast in the animation. Those slender lobes then begin to oscillate from the northeast to the southwest (orthogonal to the primary wave direction), and those oscillations create smoothly moving and growing ripples like those shown in conceptual figure 7. The rings of slender lobes begin again with each wave emergence. Droplets can be seen being stripped off of the waves at every wave depth and, especially, the thinning oscillating slender lobe ridges. Sometimes, the ridges don't simply oscillate; they form dual spreading ridges which take on that familiar RTI mushroom shape. When these occur, the alternating signs of tangential velocity prevalent in oscillating slender lobes would be non-existent. That can be seen also in the prior-referenced tangential velocity contour animations.

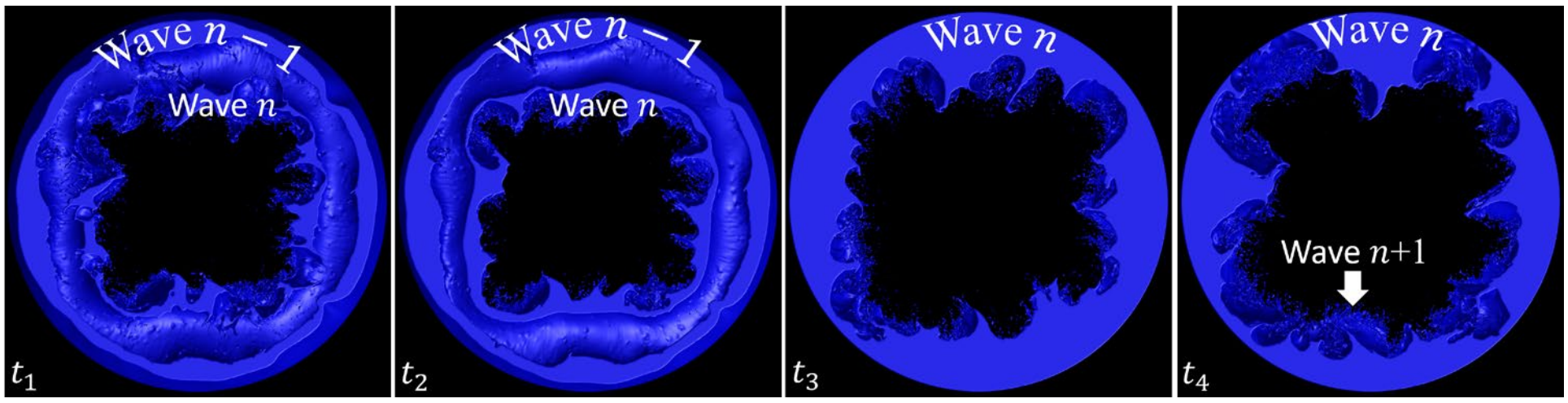

Figure 11: Instantaneous iso-surfaces of fuel volume fraction, looking from the end of the nozzle back into the flow, at four different arbitrary times progressing from left to right. A time-lapse animation of can be found here.

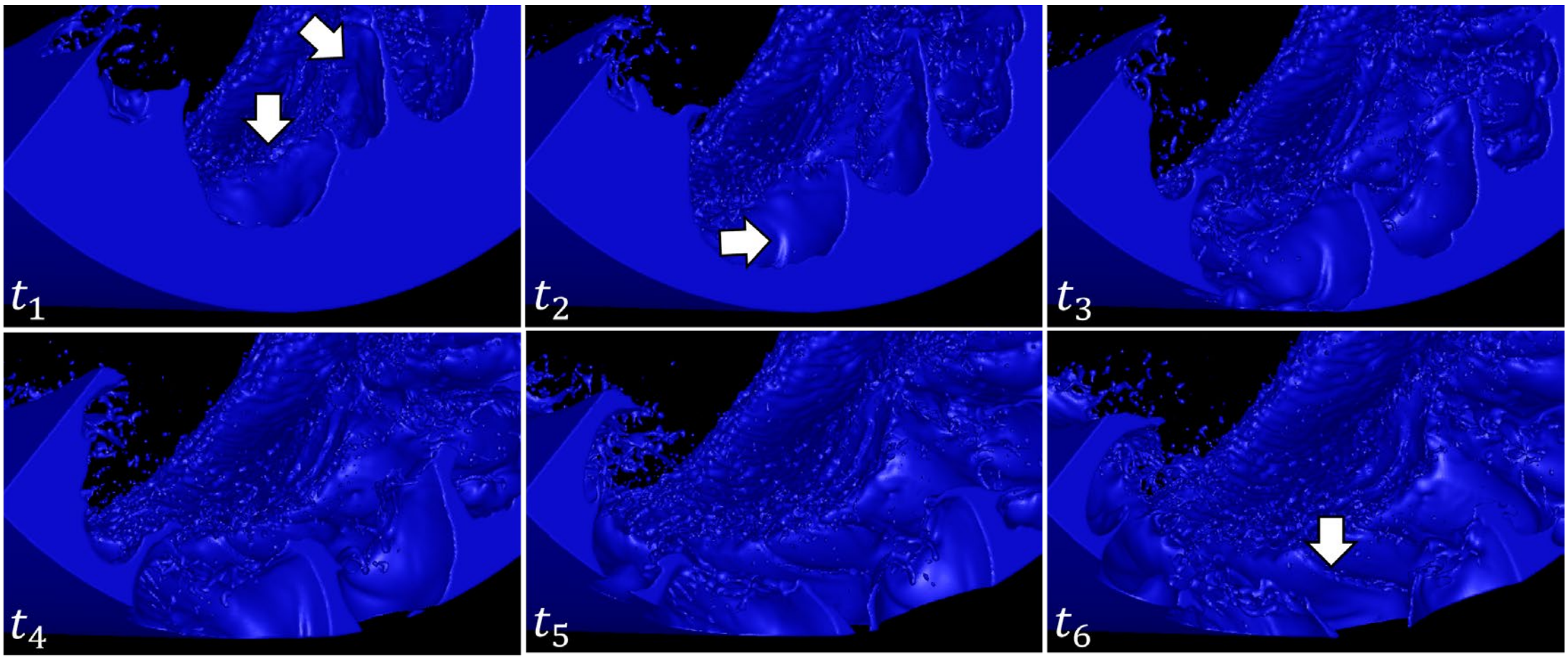

Figure 12: Instantaneous iso-surfaces of fuel volume fraction, looking obliquely from the end of the nozzle back into the flow, at six different arbitrary times progressing from left to right and then top-down. White arrows annotate the ripples which develop and then cause further azimuthal deformation. Flow is generally from northwest to southeast, and a time-lapse animation of a broader view can be found here.

### 4.4. *Consequence = SASS*

Quantification and characterization of the SASS behavior will now be sought, beginning with figure 13. Six images are provided for four different simulations showing the maximum tangential velocity magnitude on a given cross-cutting plane. Preliminary simulation data has been removed from the body of data considered here, such that only the most recent QSS time-series data are presented. The abscissa is flow time which has been normalized by dividing by the wave time scale $t_w$. (Throughout this document, the # of wave times, WT, will be used synonymously with the phrase "convective times", CT.) Any time in this figure or future figures there is no ordinate shown on images, it means that they share the same ordinate (quantity and range) with the image directly to their left. The same goes for the abscissas; any time an abscissa is not shown, it is the same as the image below it. All but image (d) cover a time span of $5t_w$, while image (d) covers $70t_w$. Image (f) is for the 20 axial locations of planes dictated by equation 4.13, but the other images are for 10 axial planes at $(x - x_0)/r_0 = 0.445 * L/D$ 1.24. (Since the latter group of planes is spaced out farther and extend beyond the high resolution mesh transition point of R3, they are not as important and are only used for relative case-by-case comparisons. Mesh R3 is the focus of this work and the only model with both the 10 and 20 plane groups; all other cases only use the 10.) The top row of the figure compares the results between the 90° low resolution mesh (70 total $t_w$ worth of data but only the last 5 shown) with the 180° low resolution mesh (40 total $t_w$ worth of data but only the last 5 shown). Image (c) can be compared to either on the top row, and it can be compared to its longer-runner data set in (d). Then, image (e, 10-plane data) can be compared to any of the previous plots or its 20-plane counterpart in (f). The R3 simulation was run more than $10t_w$ after interpolation from prior low resolution results, but only the last $5t_w$ were considered herein. The first $5t_w$ of that model's run time had to be discarded since the new, more highly resolved flow behavior took time to develop.

Beginning with images (a)-(d), all for the coarse simulations, four assertions can be made. First, tangential velocity generally decreases with axial distance and will be revisited later. Second, the velocity signals are periodic. Third, there is no obvious effect of modeled azimuthal domain size. Fourth, the longer time frame of image (d) shows sustained replication of earlier times. Image (e) for the most resolved mesh shows very similar trends, but peak magnitudes are higher for some of the planes. This indicates the somewhat diffusive character of the R1 mesh. Lastly, image (f), whose planes are more concentrated in the more tangentially active region, illustrates significantly higher swirl velocities. They occur at the same period as the lower resolution mesh; frequency content will be evaluated later.

Eventually, TA planar statistics will be presented, so we now demonstrate in figure 14 that $\approx 5t_w$ is sufficient to develop stable TA data. This figure has the same layout and flow time scales as figure 13, and similar conclusions are drawn. Model azimuthal domain does not matter much, the $70t_w$ results do not differ much from the $5t_w$ results (though flatter), mesh resolution matters some, and more planes focused in the orifice area reveal higher TA values.

Temporal standard deviation of the maximum planar tangential velocities is depicted in figure 15, the layout of which matches that of the previous two figures. In this case, however, the plotted values are the cumulative coefficients of variation (COV) of the various velocity signals. In other words, just like a cumulative TA, the ongoing standard deviation data are continually collected and normalized by a running TA. The data streams in figure 15 are much noisier than those in figure 14, as it is expected that a second moment is less stable than a first. Though they are noisier, some of the same conclusions from earlier figures are inferred. Aside from statistical variation, there is minimal effect of the model's azimuthal angle.

Figure 16 offers continued illustration of SASS behavior by providing tangential velocities at various monitoring points inserted both in the streamwise/radial and azimuthal directions. The five streamwise/radial locations are “Inner” (near the early, calm fuel-steam interface), “Middle” (near the fuel-steam interface but farther downstream), “Exit” (at the fuel-steam interface near the nozzle exit), “Tip” (farther downstream near where the wave blasts radially outward), and “Outer” (well outside the nozzle). There are five each of those at azimuthal angles of 4.5° , 13.5° , 22.5° , 31.5° , and 40.5° (with an arbitrary 0° reference angle), making 25 points total. Further point location details are in Wilson *et al.* (2023*a*). Each row in figure 16 delineates a particular point streamwise/axial location family. The LH column involves the last 5 $t_w$ for mesh R1, the middle column includes the last 5 $t_w$ for R3, and the RH column shows the last 70 $t_w$ for R1; all results are from a 360° model. This figure serves a four-fold purpose. First, there is a clearly quasi-periodic signal in the local tangential velocities, oscillating about 0; though it is far from perfectly sinuous, especially considering the RH column showing the high levels or irregularity. Second, the general magnitudes of the alternating tangential velocity can be garnered: inner $\approx \pm 0.3$, middle $\approx \pm 30$, exit $\approx$ 30, tip $\approx \pm 20$, outer $\approx \pm 10\,m\ s$. Third, it is clear that considering only 5 $t_w$ of instantaneous data can be misleading for determining temporal minima and maxima; notice the longer term data typically covers a wide range of dependent variable. Fourth, R1 data are different from R3 data. However, R1 and R3 are not any more different from one another than the shorter time R1 data are from the longer time R1 data. In other words, mesh R1 is apparently sufficient.

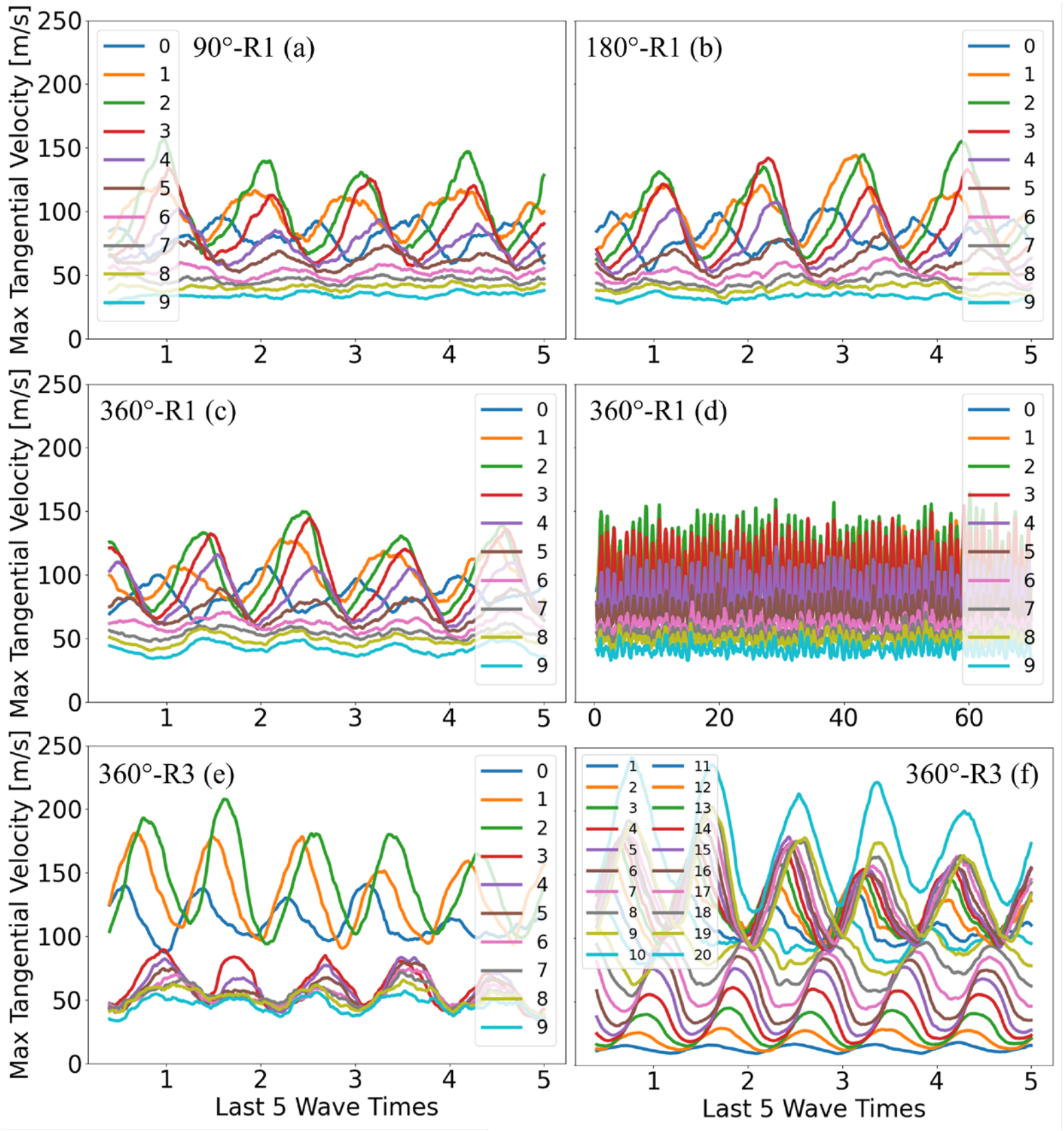

Figure 13: Maximum tangential velocity component magnitude on various planes from four simulations, each with the particular modeled azimuthal domain size and mesh resolution listed at the top of each image. Plots involving Planes 0-9 are not at the same axial locations as those described by earlier equation 4.13 and are less important. They follow a different axial pattern defined by $(x - x_0)/(r_0) = 0.445 * ID - 1.24$, and their primary purpose is to compare less resolved models to one another throughout this paper. Any time in this figure or future figures there is no ordinate shown on images, it means that they share the same ordinate (quantity and range) with the image directly to their left.The same is for the abscissas; any time an abscissa is not shown, it is the same as the image below it.

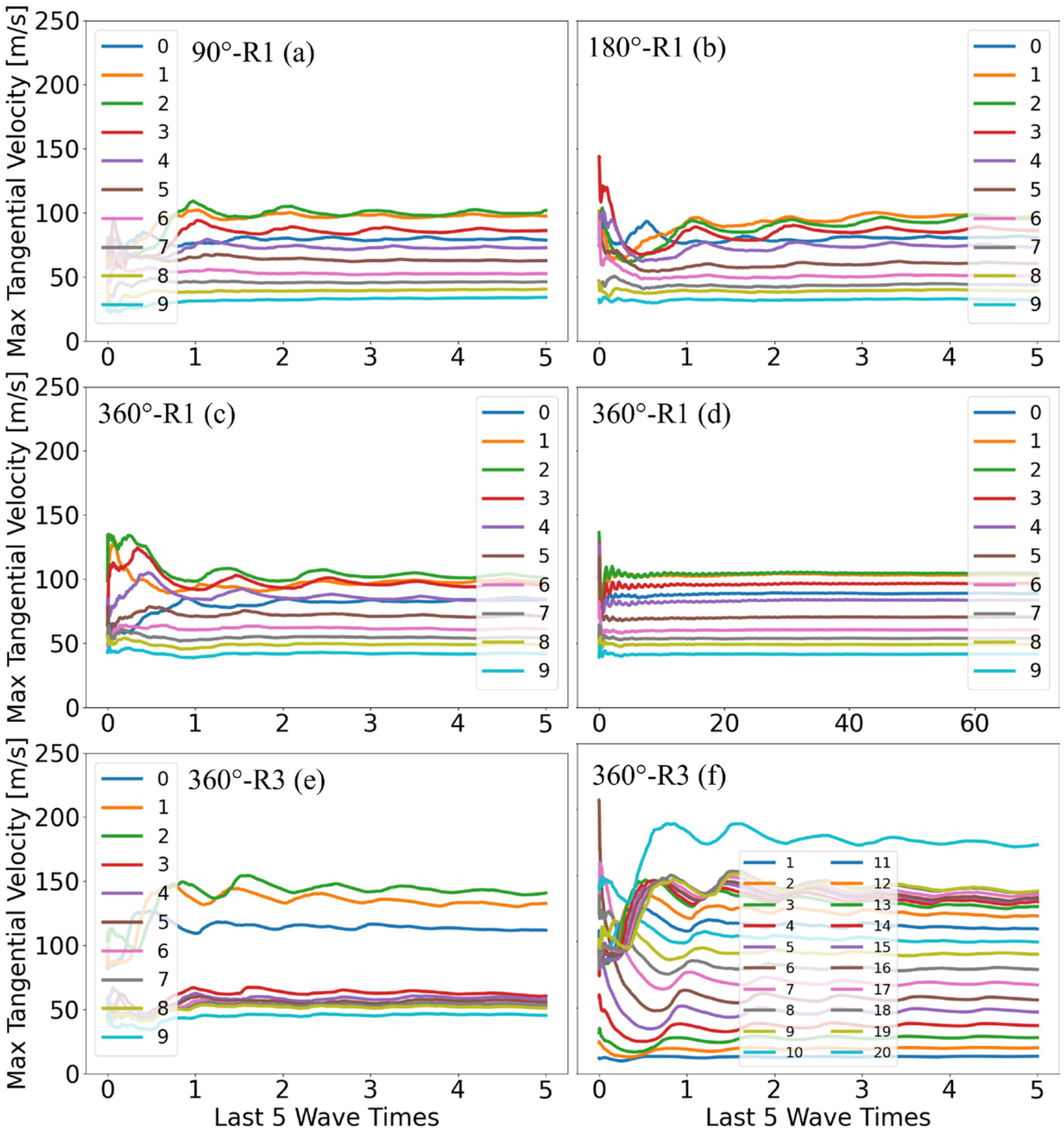

Figure 14: Cumulative TA maximum tangential velocity component magnitude on various planes from four simulations; the figure layout and flow time scales are the same as with figure 13.

Despite the strongly irregular transient character displayed in figure 16, figure 17 shows that the cumulative TA tangential velocities quickly converge to $\approx$ 0 in a relatively short period of time. The LH column shows the last 70 $t_w$ for mesh R1, and the RH column involves the last 5 $t_w$ for R3. The goal of achieving 0 as the mean is essential for two reasons. First, it confirms that all the signals are indeed oscillating (however irregularly) about a truly neutral value. In an axisymmetric geometry and purely direction-switching behavior, the baseline must be 0. Second, it means that we have enough data from which we can draw conclusions (future plots). Obviously, at longer averaging times, the cumulative TA becomes closer to 0; on the other hand, even after only 5 $t_w$ for both meshes, the data signals have *significantly* smoothed and are close to 0. In fact, each of their departures from zero is < 1% of their temporal standard deviations.

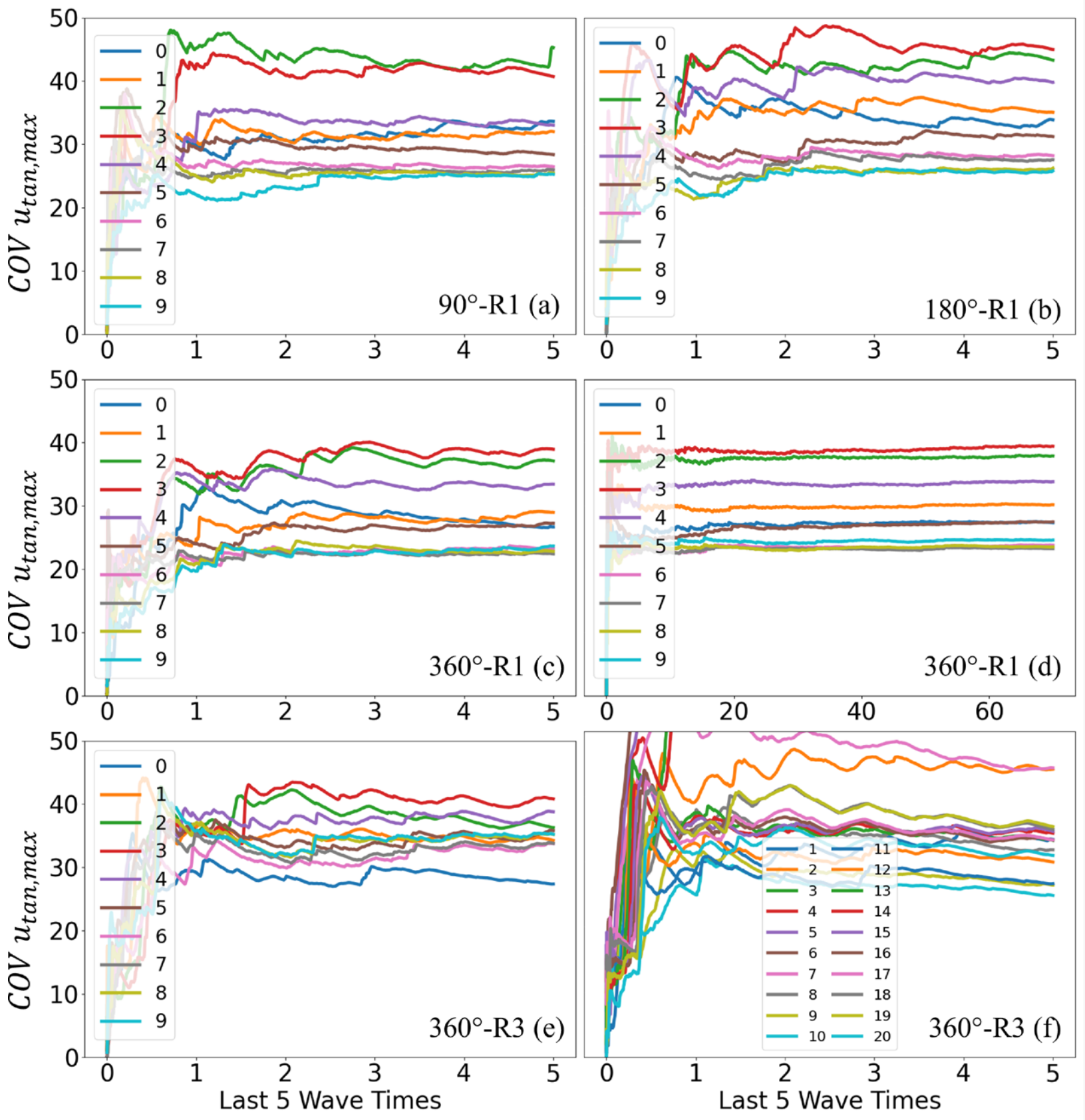

Figure 15: Cumulative coefficient of variation (COV) of maximum tangential velocity component magnitude on various planes from four simulations; the figure arrangement and flow time scales are the same as with figures 13 and 14.

Those standard deviations in the oscillating tangential velocities for various meshes and simulation run times are shown in table 4, where each value is averaged among the appropriate 5 azimuthal positions. Evidently, the exit point has the highest level of temporal variability, followed by the middle point, tip/outer points, and lastly (by far) the inner point. It may not be obvious in the raw data from the table, but there are clear trends in the standard deviation when considering the effect of simulated azimuthal angle and data collection time as shown in figure 18. All values in the LH image have been normalized by their 360°-R1 values and involve relatively large simulation data sets for the coarser meshes. It is seen that the simulated domain angle has a significant effect (up to $\approx$ 40%) on the SASS behavior and that a fully 360°mesh is needed for its assessment. The RH image represents values normalized by their respective values at the highest simulation runtime for the 360°-R1 model. This shows the strong need for excessive simulation run time to allow certain statistics for the ”tip” and ”inner” points to fully stabilize. With figure 18 in mind, it difficult to say if the differences between the 360°-R1 mesh and the 360°-R3 mesh in table 4 for those two points are a reflection of mesh resolution differences or a lack of data accumulation time for the more refined mesh. Core-hours have already been discussed, and it is implied that $>$ 100 wave times for the R3 mesh would require exa-scale computing. This assumes that a PLIC VOF model would show typical scaling behavior with core count - see Prichard & Strasser (2024) and Prichard & Strasser (2025). It also may be of interest to reflect on figure 15 again which also involves standard deviations. Those, however, are based on planar maxima and seem to converge much faster than local point-value standard deviations.

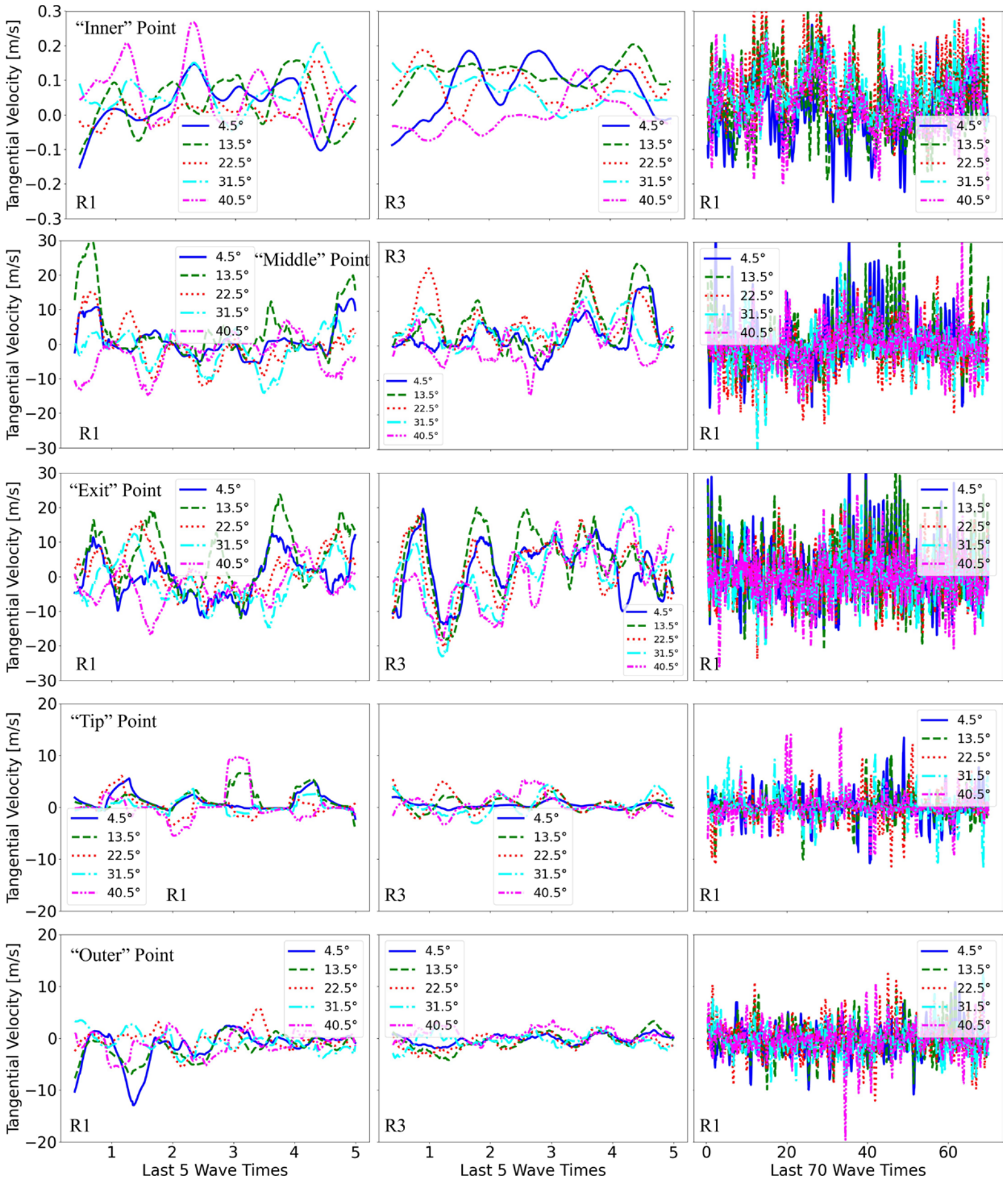

Figure 16: Tangential velocity component magnitudes on monitor point locations discussed in Wilson *et al.* (2023*a*). Different line colors are used for different azimuthal locations of those points. The LH column involves the last 5 $t_w$ for mesh 360°-R1, the middle column includes the last 5 $t_w$ for 360°-R3, and the RH column shows the last 70 $t_w$ for 360°-R1.

Further analysis of the fluctuating behavior will now be discussed. FFT results from the planar tangential velocity maxima are provided in figure 19, where the top row is for the 360°-R1 mesh at two data collection times. Circles mark peak frequency values for easy identification. The majority of the 10-group planar values peak at$\approx$ 1000$Hz$ for both data collection sizes, while the larger data set has cleaner (more focused) FFT peaks. Plane 0 in both of them seems to prefer the next harmonic. With the top row in mind, we consider the bottom row for the 360°-R3; the LH image is for the 10-plane group and RH image for the 20-plane group. The finer mesh results shown a similar trend as the coarser mesh for the same data collection size, and most planes peak at$\approx$ 1000$Hz$. The obvious correlation between axial planar location and FFT amplitude for the bottom row RH image will be visited later.

As with figure 19, a similar FFT presentation for the "middle" point is included in figure

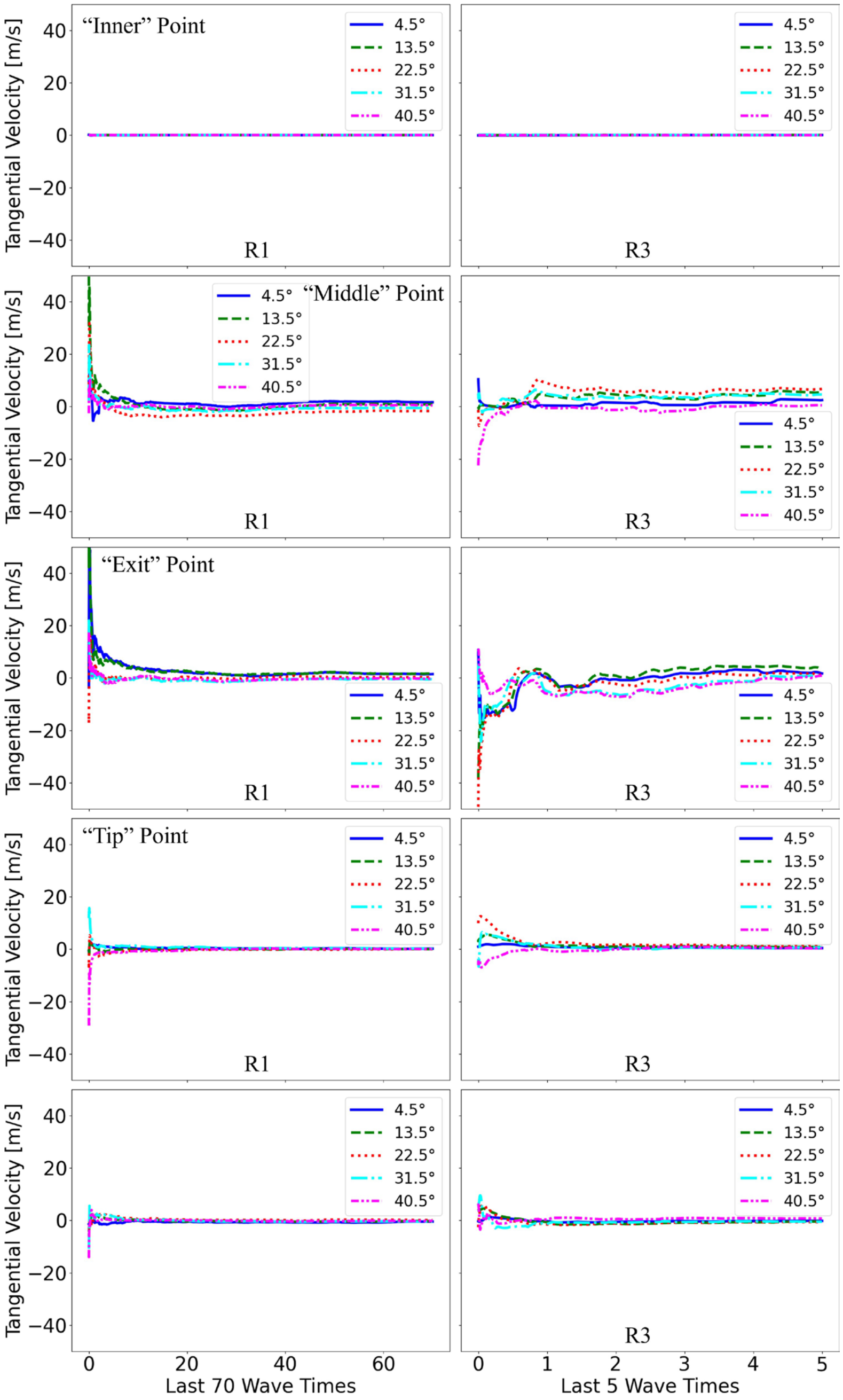

Figure 17: Tangential velocity component magnitude cumulative TA on monitor point locations discussed in Wilson *et al.* (2023*a*). Different line colors are used for different azimuthal locations of those points. The LH column involves the last 70 $t_w$ for mesh 360°-R1, and the RH column shows the last 5 $t_w$ for 360°-R3.

| Location | 90-R1 | 180-R1 | 360-R1-L5 | 360-R1-L70 | 360-R3-L5 |
|---|---|---|---|---|---|
| Inner | 10 | 11 | 10 | 10 | 6 |
| Middle | 1630 | 1540 | 1370 | 1360 | 1270 |
| Exit | 2490 | 2110 | 1790 | 1910 | 2090 |
| Tip | 800 | 674 | 596 | 559 | 276 |
| Outer | 626 | 559 | 688 | 546 | 343 |

Table 4: Tangential velocity component magnitude standard deviations (in m/s) on monitor point locations discussed in Wilson *et al.* (2023*a*). As the reader may recall, the first number is the simulated azimuthal angle, the R value is the mesh refinement code, and the $Ln$ at the end is the number of most recent $t_w$ worth of data included in the standard deviation.

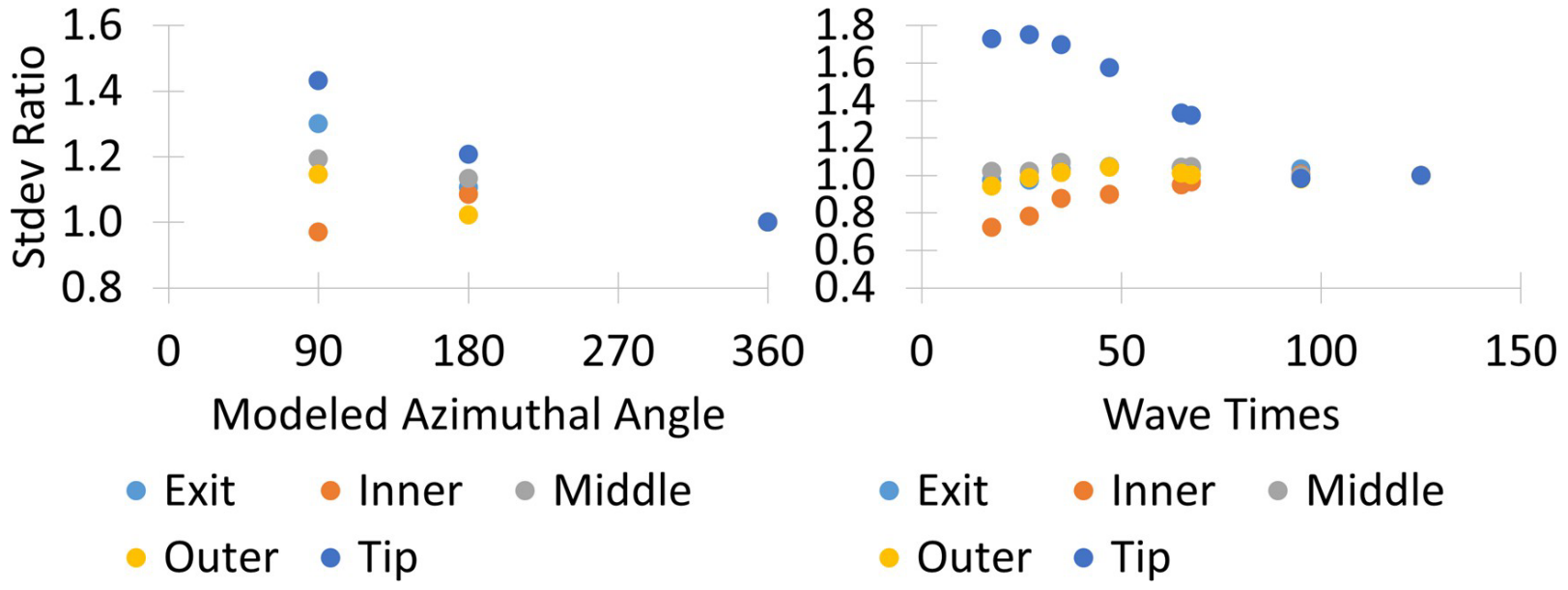

Figure 18: Normalized standard deviations of point value tangential velocity component magnitudes. In the LH image, the values of all point standard deviations have been divided by their 360°-R1 values (see table 4 for raw values) to illustrate the trending with simulation azimuthal angle. For the RH image, all point standard deviations have been divided by their longest running 360°-R1 simulation ending value to illustrate the trending convergence with data collection time.

20. Each mesh and data accumulation time is labeled in the three images. Again, the different colors represent the different azimuthal locations of the middle points. It is clear that all points generally take the peak value of $1000Hz$ for both meshes and data collection times. The low collection time plots for the two meshes (first and second from the left) are as different from one another. The lower collection time is from the higher collection time (far right) for the coarser mesh. Though the longer time data is cleaner, all points still do not share the same peak frequency and amplitude. This indicates that more than 70 $t_w$ is needed to have perfectly consistent FFT results.

The next figure 21 aims to characterize axial distance tendencies in various metrics for multiple simulations. The LH column carries the greatest weight and will be discussed first, as it displays the more finely spaced 20-plane group for the 360°-R3 simulation. TA maximum planar tangential velocity, normalized by $W_c$, versus distance from the start of the pool is given on the top-left. An approximately linear trend is seen from the start of the pool at the origin up to $\approx$ 40% at P13 and then tapers off after that. Even through the orifice region this trend is persistent. Evidently, after the CAF is no longer confined, the SASS behavior is no longer amplified. Below that image are data for the tangential velocity temporal COV versus distance from the pool beginning. Along the initial growth of the wave, upstream of the orifice, there is a growing COV versus distance up to more than 60%. In the orifice and beyond, the COV is relatively flat in the 30% range. The next lower image is the dominant FFT frequency, $f$, versus distance, revealing values on the order of $1000Hz$ through most of the system; the exception appears to be an outlier of the next harmonic just after the end of the orifice. This region is particularly difficult to characterize due to the fact that the CAF is just becoming unconfined. Finally, at the bottom, lies the FFT amplitude associated with $f$ . As with the COV, we see a sharp rise from the origin with a nearly linear trend up to ≈ 30% just upstream of the orifice and a sharp decline within the orifice. Unlike the COV, on the other hand, the amplitude takes on a different linear nearly trend downstream of the orifice and reaches values > 50.

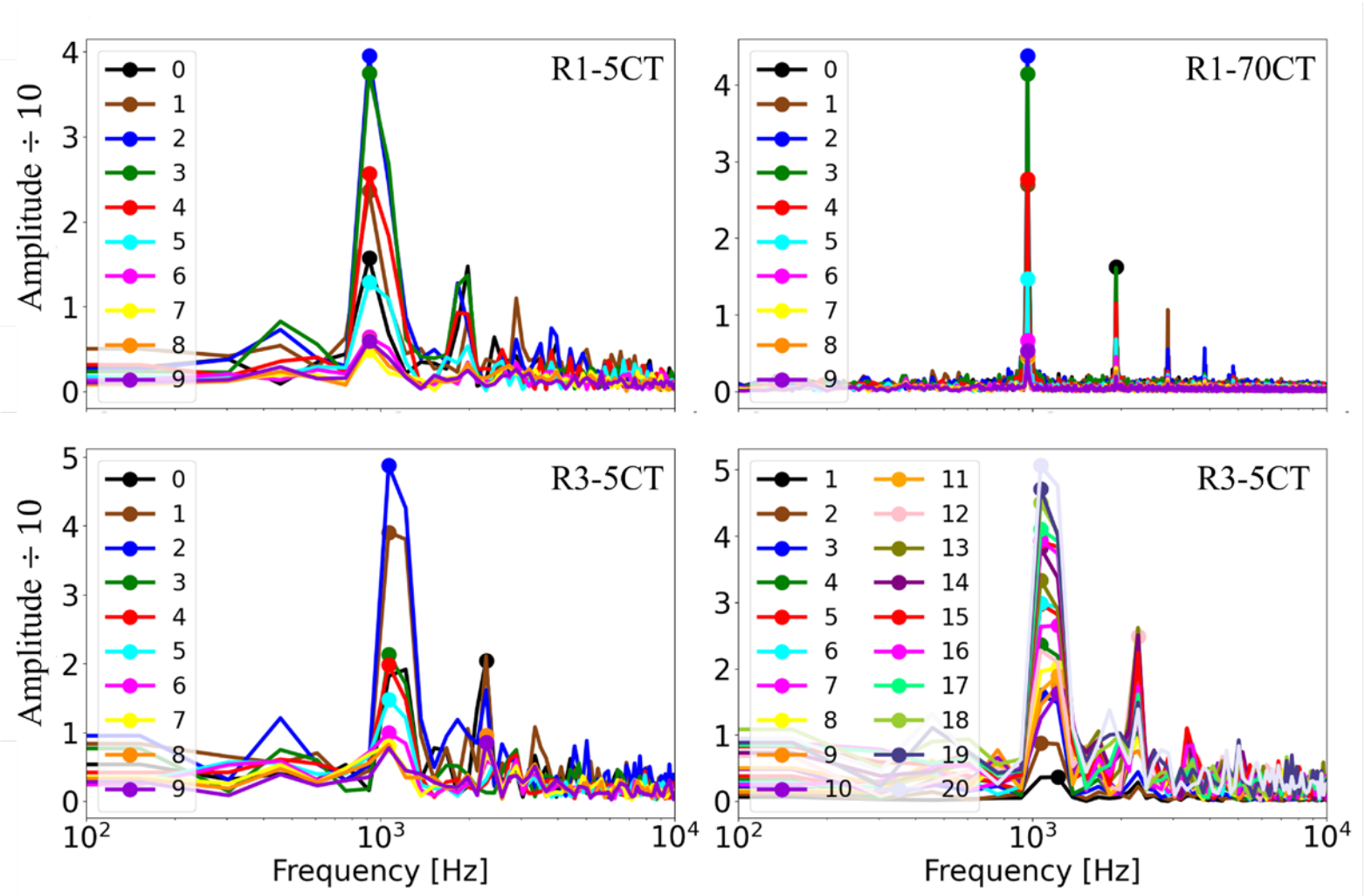

Figure 19: FFT results from various planar tangential velocities. The 10-plane group from the coarser mesh 360°-R1 at two different data accumulation times is included in the top row, while both planar groups are included for the finer 360°mesh on the bottom row. Circles are used to identify the peak frequencies.

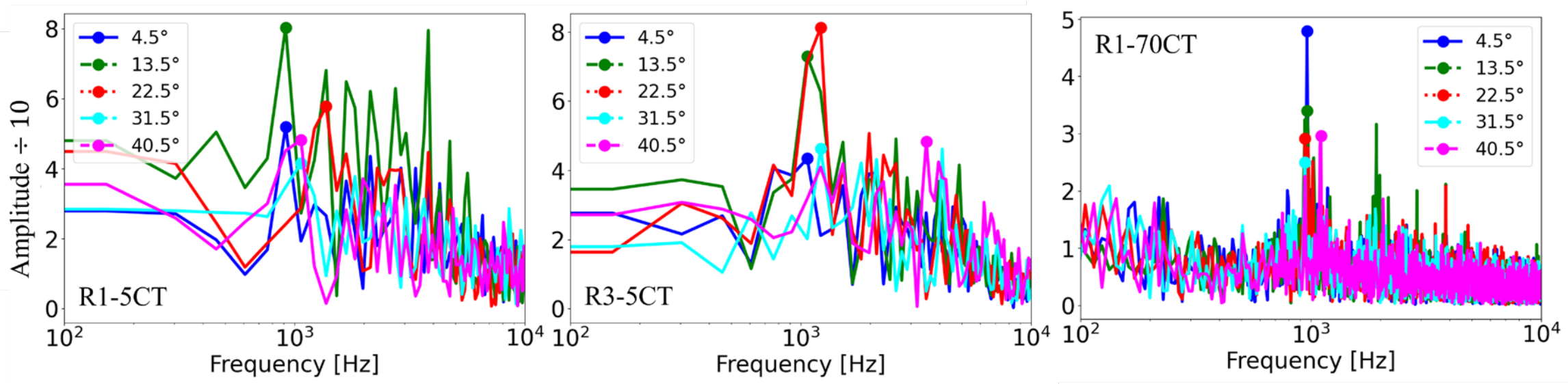

Figure 20: FFT results at the ”middle” point. From left to right, the images are for the coarse mesh at a low data collection time, fine mesh and a low collection time, and the coarse mesh at a high collection time. Circles are used to identify the peak frequencies.

The middle column entails data from both the coarse and fine 360°meshes and with the coarse mesh at two data accumulation times. This column is based on the less finely spaced and broader reaching 10-plane data collection method. Those planes start at about 0.5× $D_O$ from the pool (within the orifice region) and continue to downstream of 2.5× $D_O$. Again, the reason for the different sets of planes is that the earlier models before 360°-R3 did not include the 20-plane set and cannot be re-run to collect all of those data now. Some of the planes in the LH column overlap with those in the middle column, and so do the data for mesh R3. The consideration of multiple sets of data allows multiple conclusions. We will first address the behavior of R3 at higher distances, keeping in mind that the mesh transitions from R3 to the R1 level of resolution beyond $x/D_O$   1. From the top image, it is clear that there is a sudden drop in peak tangential velocity out beyond $x/D_O$   1. We postulate that this is not entirely due to mesh resolution changes but is probably driven by that. Next, it is seen that the COV increases again and then levels off. Then, $f$ remains near $1000Hz$ and then suddenly jumps to the next harmonic towards the end of the sampling region. Finally, FFT amplitude falls suddenly and then approaches 0 again somewhat asymptotically. Other observations can also be made from the center column of images. For R1, the two sampling data sets shows nearly identical responses, indicating that 5 WT is sufficient for sampling. The only outstanding difference is the value of $f$ close to $x/D_O = 0.5$ being higher for more sampling time. Additionally, we see significant differences between R1 and R3. The tangential velocity, again, drops very sharply where the mesh coarsens, while the coarser meshes are more gradual. They also exhibit lower values early on. The COV follows the same trend between the two meshes, but they are lower for R1. The frequencies are similar between the meshes except for the aforementioned sudden jump in R3. Lastly, the FFT amplitudes of the two meshes follow each other relatively closely.

The three R1 mesh simulated azimuthal angles are compared at their largest sampling times in the far RH column. While none of the curves appear to have overwhelming differences between the three results, a few trends emerge. The 360° mesh has the highest peak tangential velocities, lowest COVs, and highest FFT amplitudes. The effect of modeled azimuthal angle on $f$ is less consistent. Certain meshes at particular locations tend to respond at higher harmonics, but this is consistent with other FFT results presented herein showing substantial energy at $\approx 2000Hz$ and higher. Therefore, we infer that FFT results are sensitive to resolution, angle, and location.

Figures 22 and 23 seek further confirmation of the proposed mechanism and will conclude the discussion of results. They are based on the methods developed in Wilson *et al.* (2023*a*) which will be briefly summarized again here. A python script was employed to generate spatial and temporal cross-correlations of planar time-series peak local tangential velocity data sets. To do so, a reference data set was chosen to hold constant; in our case, it was the first plane of each planar groups, separately. All other columns of planar data were then compared to the reference data column, and spatial correlation coefficients were calculated. Next, each planar data column was shifted one time-step size down, and coefficients were re-calculated. This was continued until all available time-step shifts in the data were visited. After this, the *maximum* of those correlation coefficients (with respect to the reference plane) for all planes were plotted in figure 22. Finally, the time shift required to achieve that *maximum* correlation coefficient (in number of $t_w$) was documented in figure 23. The time-shifting required to reach the maximum correlation can be thought of as wave communication time, i.e. how much time is required for a laterally buffeted slender lobe tip wave to travel along the base axial waves shown in prior publications. On the abscissa of both figures, the planar reference axial location is the end of the nozzle orifice.

Beginning with the top row of figure 22, a strong correlation (between 70 and 80%) is evident in the R1 simulation across the 10-plane sampling region. This is true regardless of the accumulated data size. The bottom row, LH image shows a similar trend for R3 with an even higher correlation. For the 20-plane R3 data set (bottom RH corner), the trend is similar; there is a mild, approximately linear decline in correlation with distance from the starting plane. The existence of strong correlation in slender lobe tangential velocities confirms the existence of the proposed axially traveling wave-like pattern of the tangential velocity in the newly formed slender lobes. Lack of correlation would indicate localized spurious tangential motion bursts that may or may not be in a wave pattern.

Arranged in a similar fashion as figure 22 is figure 23. Irrespective of the data accumulation time, there is a steady rise in wave travel time and then a steadying off at about $0.5t_w$ for mesh R1 10-plane data set. The inflection point in the middle of the data, leading to a non-intuitive decline in slender lobe wave communication time farther from the reference point, may have two causes. The first is noisy data; as seen in prior figures, the azimuthal velocities are highly variable. It could be that correlation is more difficult to find farther from the orifice. The second option is that the result is physically based. For this to be true, there would need to be a sub-surface wave amplifier which bypassed planes closer to the orifice and then arose in later planes. This sub-surface amplifier would be needed to communicate information from earlier perturbations to more distant positions more quickly than it does to closer positions. The R3 mesh 10-plane data set shows a similar trend as R1 but with lower communication times. More peculiar are the 20-plane R3 results in the lower RH corner. The first pronounced feature is the stair-stepped ascent in the communication time from the pool start to the orifice. The same thing said about the inflection points for R1 can be applied to this. Another feature is the higher communication time that is reached (1.2 versus 0.4), implying a slower slender lobe wave farther upstream in the orifice than downstream of the orifice. Finally, the sudden drop in communication time after the orifice is unexpected. As with prior directional changes, we are not sure why this occurs. This brings into view the differences between the lowermost (R3) figures. The reference plane for the LH image is the first in the 10-plane series, while the reference plane for the RH image is the first in the 20-plane series. In the axial region where these curves overlap ($0 < x/D_O < 0.5$), very different shapes are observed. In the LH image, as discussed, there is a steady rise and tapering off. In the RH image, there is a sudden drop. Indications are that slender lobe wave behavior upstream of the orifice is different from that downstream of the orifice. The determination of the root cause of this is relegated to a future effort.

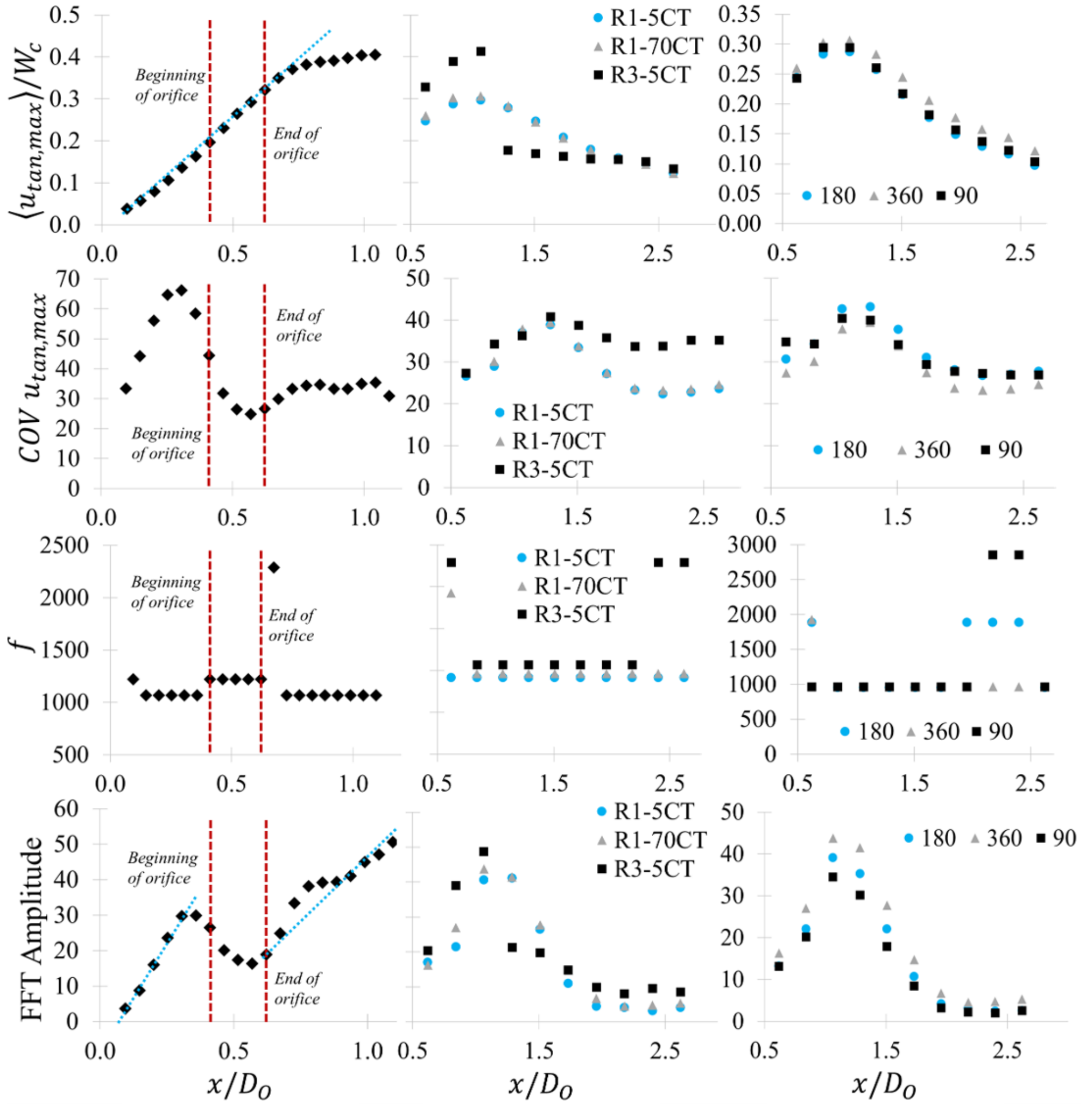

Figure 21: Time-averaged planar data for various metrics and simulations. The top row is for the maximum planar tangential velocity component , 2nd row for its COV, 3rd for the dominant FFT frequency, and the bottom for the FFT amplitude at peak $f$ . LH column details 20-plane results from 360°-R3, middle column shows 10-plane data from both 360°meshes at various data accumulation times, and the right column shows 10-plane results for all three R1 meshes at their maximum data accumulation times.

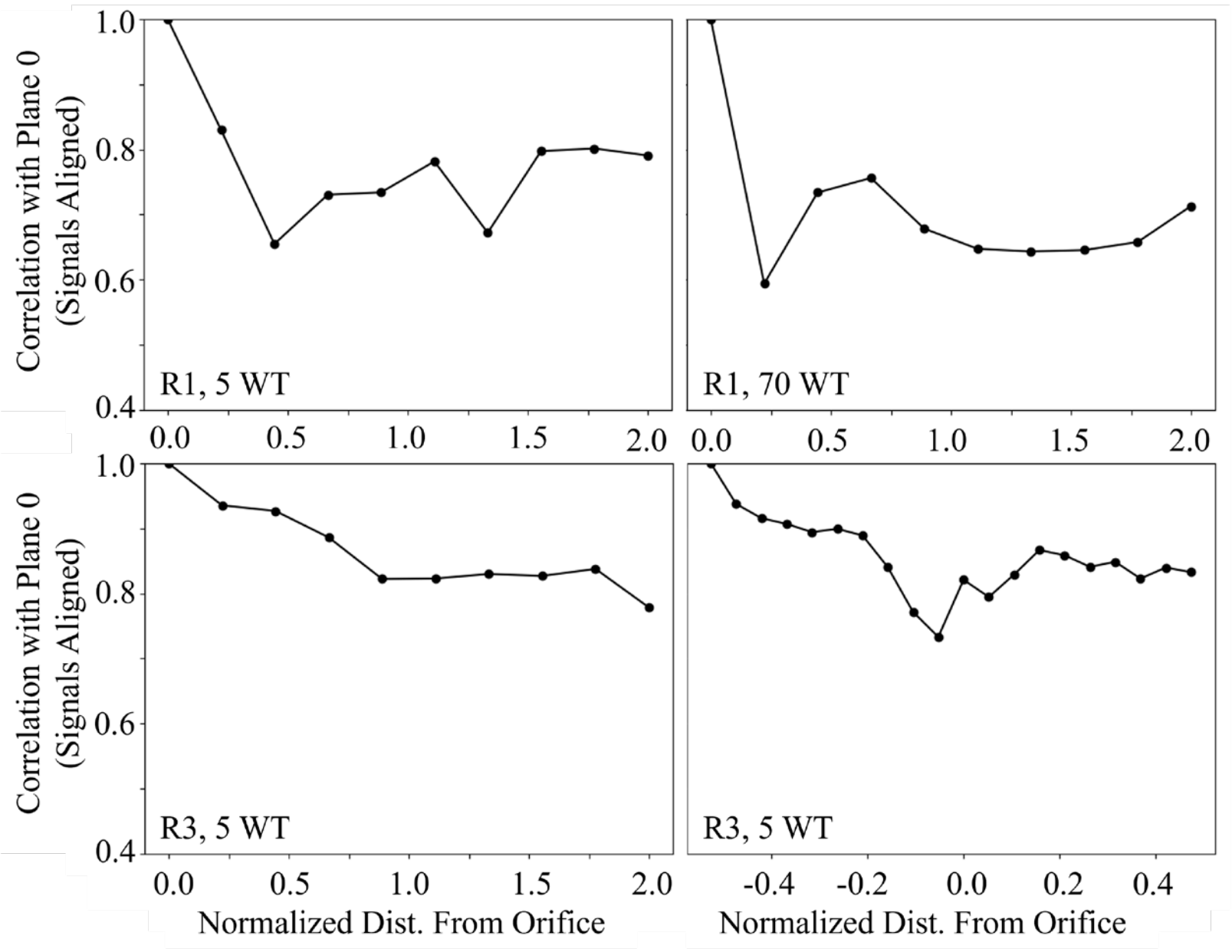

Figure 22: Spatial cross-correlations of peak planar tangential velocities for two mesh resolutions, R1 at two data accumulation times, and R3 for two different plane sets. In all cases, the reference plane is the first one in the series, and the correlation is the maximum value possible through time-shifting. Unlike prior planar plots, the abscissa is based on the axial distance downstream of the nozzle orifice.

## 5. Conclusions

A detailed computational analysis has been conducted to characterize and provide a mechanism for spontaneously axis-switching swirling (SASS) behavior in a CAF involving transonic steam and a non-Newtonian fuel. While the authors' past work explored the atomizing character of this CAF, the existence of SASS, its consequences, and its cause have never been considered. By highlighting a myriad of dimensionless numbers across the multiphase physics field, it was demonstrated that this CAF represents a significant departure from prior CAF publications. For example, unlike previously found "corkscrew" CAF patterns, SASS does not involve gravity, and waves spontaneously change directions about a zero tangential velocity. Swirling was found to be self-generated, since it was not caused by boundary conditions or atomizer geometry.. Frequency analyses revealed that the dominant frequency of this pattern is close to that of the base axial wave pattern which is$\approx$ $1000Hz$. Strong spatial correlations along the nozzle axis demonstrated a wave-like pattern in the switching tangential motion.

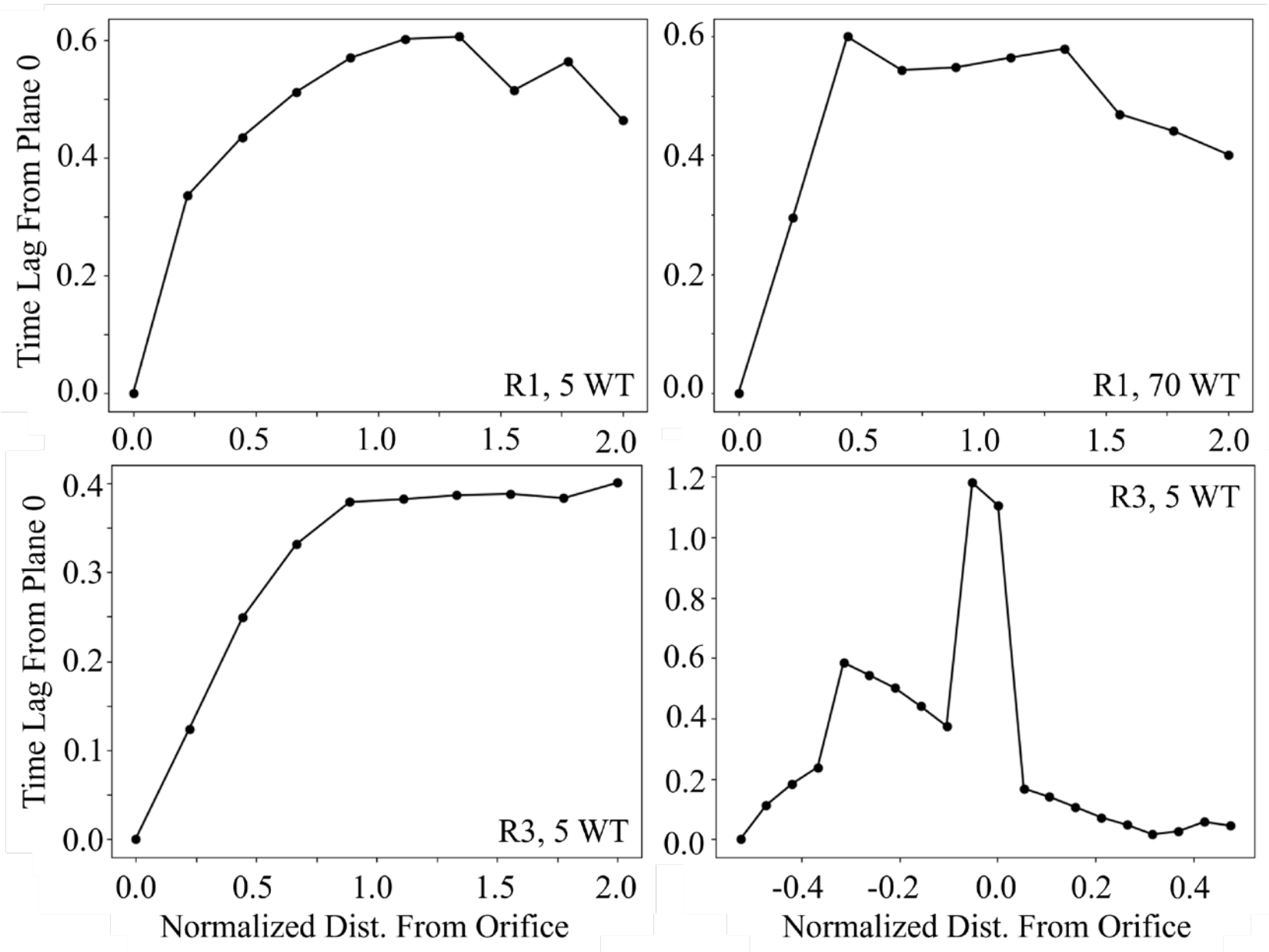

Figure 23: Temporal cross-correlations of peak planar tangential velocities for two mesh resolutions, R1 at two data accumulation times, and R3 for two different plane sets. In all cases, the reference plane is the first one in the series, and the correlation time-shift (in units of WT) is that required to achieve the maximum possible correlation shown in figure 22.

SASS resulted in strong, nearly periodic fluctuations in tangential velocities which, in a TA sense, exceeded 40% of the steam TA core velocity; instantaneous values were much higher. The rise in TA maximum tangential velocity versus normalized distance from the start of the fuel pool to the end of the orifice was almost perfectly linear with a slope of 0.554; the peak occurred at about 1 orifice diameter from the start of the pool. After the peak, it fell approximately linearly to the 10% range near the end of the modeled domain. Tangential velocity standard deviations (in m/s and rounded to two significant digits) were as high as 1300 and 2100 at the middle and exit point locations, respectively. The amplitude of the dominant tangential velocity frequency versus dimensionless distance showed a nearly perfectly linear rise from the pool start to the orifice with a slope of 127, and it peaked at 30. It fell again across the orifice. Downstream the orifice, on the other hand, the amplitude rose linearly again with a slope of 62.6 up to a value of 51. After this peak at about 1 orifice diameter downstream of the pool start, it asymptotically fell to nearly zero. COVs of the maximum tangential velocity peaked at nearly 70% just upstream of the orifice beginning, then dropped suddenly, and then leveled out to approximately 30% after the orifice.

SASS was found to be the result of pile-up of sequential base axial waves which produced radial modal deformation flow patterns. Ensuing radial pressure gradients created 3-D deformations in the base wave surface. The deformations became slender lobes and were further exposed to velocity and density gradients and the consequential instabilities. This produced a wave pattern orthogonal to the base wave, and that secondary wave wind-up precipitated the tangentially fluctuating velocities.

Although the simulation tool had been previously validated, and just like experiments have error bars, it is important with any numerical study to consider numerical uncertainty. This was thoroughly explored herein, as we varied mesh resolution, axial sampling locations, the azimuthal angle/extent of the modeled domain, and statistical data collection time. Varied responses were found for all tests, such that either extreme was sufficient for certain metrics. For example, a significantly coarser mesh revealed the same SASS behavior as the finer mesh, but some metrics were diffused by the coarser mesh. Even though lower azimuthal angle models produced the same SASS behavior and similar TA results, they were demonstrated to be inferior. Some metrics were not sensitive to data accumulation time over the range of 5 to 70 $t_w$, while others were. This was especially important given the cost-prohibitive aspect of running the finest simulation (at 152 million cells) beyond the $10t_w$ it was run for this paper. Results should be interpreted in light of the discussed sensitivity. Despite limitations, SASS behavior was clearly demonstrated and characterized herein and can be done so with a quite limited mesh.

It was clearly shown that a wide range of mesh resolution produced very similar effects, so SASS is very likely not solely a consequence of numerical instabilities. Despite the rigor demonstrated herein and in past validation works, the existence of numerical uncertainty is a given. In particular, use of the SST linear eddy-viscosity model could be detrimental to accuracy. The authors recognize that the existence of SASS has only been numerically demonstrated at this point. This implies the possibility that some of the response shown herein is numerically sourced. We highly recommend that a future endeavor include experimental validation with high-speed camera functionality along with pressure fluctuation data acquisition.

These findings have implications in atomizer design, turbulence modulation, and flow control, all of which have overlap. The aim of the optimum atomizer design is to produce the largest fuel interfacial surface area-per-volume for the least entropy generation through friction and mixing. Alternatively, one could seek to minimize pumping power. In both cases, the goal is to bend, stretch, and otherwise deform the liquid interface with as little effort as possible. This is accomplished by exciting or amplifying natural instabilities. Take turbulence modulation, for instance. Turbulence already involves the production of rotating motion at a wide range of length scales. Above a transitional Reynolds number, as is well-known, the flow is susceptible to perturbations. With SASS, there is a secondary fluctuating energy production mechanism as the fuel interface begins to rapidly switch directions while being sheared by a high-velocity steam. This further injects energy into the already fluctuating flow field. In addition to this turbulence augmentation, there is also a modulation in the sense that the primary direction of vorticity is re-oriented. For primary waves, the vorticity is aimed in the azimuthal direction. With SASS, in contrast, vorticity is produced in vortex stretching along the primary flow direction underneath the alternating fuel slender lobes; vorticity is then aimed axially. Fuel droplets are more likely to develop as a result of augmented and anisotropic turbulence, as the conversion of ligaments into smaller droplets requires shear and deformation in all three dimensions.

Since the SASS is a consequence of deep (relative to the radius of curvature) annular waves in a cylindrical channel, a designer should consider the effects of reducing the flow channel diameter, further encouraging the fuel-steam interface buckling phenomenon. Alternatively, the fuel depth could be increased by increasing the fuel mass flow or decreasing the steam mass flow. Due to finite resources, no other boundary conditions or geometries could be considered herein. Flow control is an integral part of the system, also, as the primary pulsing waves are modulated by SASS motions of a comparable frequency. These pressure and flow rate disturbances feed back onto the inlet flows of both the steam and fuel; the resulting pressure pulsations and upstream instabilities were discussed in prior works. It would be helpful to consider alternate designs and boundary conditions in a future effort.

Finally, more quantification may reveal interesting features of SASS. Pressure gradients have been touched on briefly in Wilson & Strasser (2022*b*), Wilson & Strasser (2022*a*), Wilson *et al.* (2023*a*), and Wilson *et al.* (2023*b*), but there has been no quantification of the radial pressure gradients driving the buckling in SASS. With respect to vorticity, having it available in cross-sectional planes (see Strasser & Battaglia (2018)for a different self-pulsing atomizer) to supplement the shear rate given in our figure and animations could be instructive. A future effort could also include investigating energy spectra like the authors' previous work in Strasser & Battaglia (2017).

**Acknowledgements.** We gratefully acknowledge the support of Liberty University's School of Engineering Dean, Ken Camplin.

**Funding.** This work was funded internally by Liberty University.

**Declaration of interests.** The authors report no conflict of interest.

## 6. Appendix

### 6.1. *Radial Flow Deformation, Broadly Speaking*

It may be helpful for the reader to appreciate the broad interest in radially modal deforming patterns, sometimes referred to as "dendritic". Related to symmetry-breaking instabilities, they appear in an astronomically wide range of situations. These include art (Canabal *et al.* (2020)), water drainage systems (Radko (2023)), crystal formation (Oikawa & Kurita (2016) and Kumar *et al.* (2020)), bacterial colony organization (Si *et al.* (2018)), surface waves in a cylindrical tank (Shao *et al.* (2021)), neuron formation (Sugimura *et al.* (2007)), healthy and diseased lung muscle development (Kato *et al.* (2022)), epithelium cell growth (Trushko *et al.* (2020) and Nelson (2016)), cortical tissue development (Garcia *et al.* (2018)), viscous fingering (Brulin *et al.* (2020), Suzuki *et al.* (2020), Sharma *et al.* (2021) and Moffatt *et al.* (2021)). Of particular interest are the images in Sharma *et al.* (2021), where they studied single-phase porous media flow of miscible fluids; their annular ring was a more viscous fluid. They found these flowing patterns above a "mobility ratio" (natural log of fluids' viscosity ratio) of $O$(1 . The actual value of the critical mobility ratio depended on the competition between convection and diffusion. Additional applications include solidification of metal droplets (Gielen *et al.* (2020)), wildfire fronts (Harris & McDonald (2022)), self-organization of micro-swimmers (Ramamonjy *et al.* (2023)), and clover shell modes in cylindrical metal buckling (Turzi (2020)). According to Nelson (2016), the wavelength $\lambda$ formed during soft matter buckling is a function of the film thickness and material properties and that the wrinkle formation does not require an external force; in-plane stresses can lead to folding. Anti-symmetric wrinkling is apparently the most energetically favorable.

These radial modal deformations, generically speaking, can be described mathematically a few ways, and figure 24 illustrates two of those ways on generic $x$ and $y$ axes of arbitrary scaling. The green curve is called a "quatrefoil", "four-leaved rose", or a "quadrifolium" and is given by equation 6.1. The black curve is in a class of "Bernoulli's Lemniscate" and is given by equation 6.2.

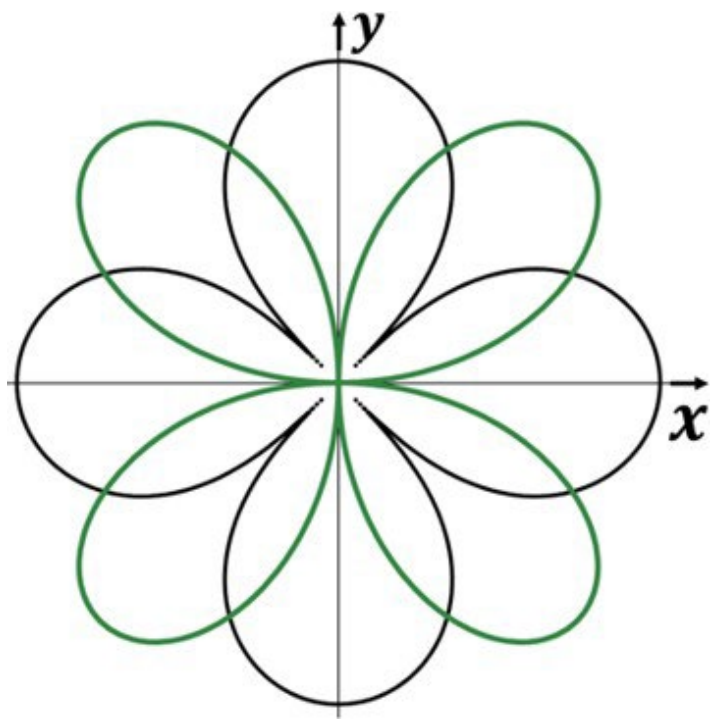

Figure 24: Sample curves on generic $x$ (horizontal) and $y$ (vertical) axes of arbitrary scalings. They are defined by equations 6.1 (green curve) and 6.2 (black curve).

$$(x^2 + y^2)^3 = cx^2y^2 \tag{6.1}$$

$$(x^2 + y^2)^{28} = c(x^2 - y^2)^{14} \tag{6.2}$$

In both cases, $c$ is some irrelevant constant. Both fall into the category of "roses" and can have various lobe counts based on exponents. For example, the exponents of equation 6.2 don't have to be 28 and 14, but there are a limited number of *even* powers which produce 4 lobes. Some exponents produce asymmetrical shapes. The "rose" family is similar to the "sinusoidal spiral" family of curves which can also vary in lobe count based on exponents. Clearly, the LH instantaneous Mach number contour plot in figure 2 is more complex than the current image and cannot be directly described by either of these curves.

### 6.2. *Core-Annular Flow*

Studied since as far back as at least the 1960s, rich fluid behavior results from interfacial instabilities when materials of very different properties share flow area in a pipe (Hall Taylor *et al.* (1963)). Even before that, however Hanratty & Engen (1957) began studying ripples on a liquid surface with a co-flowing turbulent gas. The discussion of CAF employs many potential dimensionless numbers; facilitation of that discussion in the context of the current research is provided by tables 1 through 5 based on data from Wilson & Strasser (2022*b*), Wilson & Strasser (2022*a*), Wilson *et al.* (2023*a*), and Wilson *et al.* (2023*b*). CAF also involves some nuanced terms from the original authors in cited papers; the reader is encouraged to see those original papers if the terms are confusing. We only include discussion of CAF flows which begin as *concentric*, though *eccentric* CAF flows exist. The four literature-based "surface tension parameters" in table 5; it is interesting that they span 11 orders of magnitude for a single flow scenario. Given the high Weber number of the current work (see table3), we take the position that, despite the wide range of values in surface tension parameters, surface tension is not a controlling force herein. "Lubrication" flow, where a more viscous oil core is surrounded by a less viscous lubricating layer of water, was studied by Miesen *et al.* (1992) at viscosity ratios of at least $3,000$ and $\beta$ up to 0.1, and $Re_{2,AA}$ of $2,000$. The fluids had equal densities, and gravity effects were considered. They found that axisymmetric modes were always the most unstable, so they confined their study to 2-D. Their wave speed was about equal to the core speed. Their growth rate was nearly inversely proportional to viscosity ratio; surface tension slowed growth but only for small wavelengths. Like our past work, they maximum growth rate was at about $\alpha = 0.7$, and their $\Lambda$ was $O(10$ .

Kerchman (1995) studied the opposite; the core fluid was more viscous than the annular fluid (like the present research). They were limited to flows of $\beta << 1$, equal fluid densities, low Reynolds numbers, and $S \sim O(1$ . Instabilities produced collar-like structures leading to a wave train and coalescence of those waves with patterns strongly dictated by $S$. They proposed that, since their surface tension was close to the value of $\beta$, capillary pressure induced by interfacial distortions dictated the order of magnitude of wave velocities. Later, Bai *et al.* (1996), studying a similarly more viscous core with fluids of matching densities, proposed that the viscous core moved like a solid. Axisymmetric bamboo waves resulted, along with higher pressures at wave fronts. Although their flows were laminar, they found decreasing wavelengths for increasing Reynolds numbers.

The first record of non-axisymmetric patterns in gravity-affected flow the authors found was Renardy (1997). Waves traveling in the azimuthal direction (corkscrew waves in a periodic orbit) or standing waves ("snake" waves) were demonstrated. Both wave sets operated in the axial direction; snake waves were simply translated along the axial dimension while corkscrew waves traveled in both the axial and azimuthal directions. Under certain conditions, they found bifurcations between the two types. When the annular fluid was the less viscous of the two, the preferred pattern was corkscrew for narrow annuli and snakes for wide annuli. In all cases, they studied relatively low values of $J$, $\zeta$, $K$, and Reynolds numbers.

According to the review paper of Joseph *et al.* (1997), perfect CAF is only possible for vertical flows or horizontal flows in which the two fluids have a common density. When more viscous flows are in the annulus, the flow is always unstable. Lubricating flow can be stable to both axisymmetric and non-axisymmetric disturbances. Only thin lubricating layers are typically stable. They say that inertia has a stabilizing effect, but they also say that increases in the Reynolds number results in an instability due to interfacial friction which is proportional to the viscosity difference. Generally speaking, for vertical flows where the core fluid is lighter (buoyancy and the pressure gradient act in the same direction for upflow), core stretching produced bamboo waves and core modal deformation will buckle into corkscrew waves.

An atomizer was investigated in Chen & Li (1999), where a compressible gas stream was in the annulus, and was shown to have 3-D cylindrical buckling-like shell mode deformations by varicose and sinuous instabilities. They demonstrated that gas compressibility became relevant in a dispersion relationship via the normal stress boundary condition between the phases and that 2-D modes were less stable than 3-D. In situations with unity $Ma_f$, the liquid was always unstable to all disturbance wavelengths due to the fact that pressure perturbations at the interface are always out of phase with wave motion. Therefore, gas phase compressibility promotes disturbance growth, while raising $Z$ lowered growth rates. At $Z \sim 1$, only at very high disturbance wavelengths did 3-D modes grow at all. At other $Ma_f$, upper and lower cutoff wavenumbers existed due to the stabilizing effect of surface tension. The reciprocal of the resulting droplet size was proportional to the most unstable wavenumber.

Kouris & Tsamopoulos (2002) evaluated a scenario of the lower viscosity fluid was in the core with a priori limitless annular film thickness, amplitude of the evolving disturbance, and domain length. It was determined that high phase viscosity gradients at low liquid Reynolds numbers were responsible for organized wave trains versus chaotic disturbances and that the wave velocity was related to the velocity difference. This makes perfect sense, given that viscosity hinder shear rate, creating more drag on the annular phase. Above $m > 1$, however, viscosity difference did not strongly affect wave shape. They showed saw tooth waves at higher $m$ and bamboo waves otherwise. Additionally, opposite the effects of Weber and Reynolds numbers, $m$ did not control wave amplitude. Varying the inverse Weber number by more than three orders of magnitude increased wave amplitudes proportionately, but wave period increased with the inverse Weber number logarithmically.

A viscous liquid annular jet surrounded by inviscid motionless gas influenced by three-dimensional disturbances was considered by Chen *et al.* (2003). They discovered that axisymmetric modes were the most unstable. They did so by using a linear stability analysis of small disturbances to evaluate asymptotic cases of the important governing parameters such as Reynolds number, $m$, and sheet thickness. Like our work, they demonstrated azimuthal waves, which are indicative of 3-D liquid shell distortion.

The relative effects of buoyancy and hydrodynamic forces in horizontal lubricating CAF was evaluated by Ooms *et al.* (2007). Their aim appeared to be determining over what range of a balance of these two forces produced concentric flow. As expected, they showed an increased eccentricity with decreasing axial pressure gradients or higher phase density differences. In the same year, Selvam *et al.* (2007) used an eigenvalue analysis, along with quantifying energy source contributions, to study vertical CAF with modest viscosity differences (including a more viscous core and less viscous core) and no density difference. Immiscible fluids without interfacial tension were found to be unstable at any $m$. A Reynolds stress term emerged as the primary source of energy production for both corkscrew and axisymmetric. At higher Schmidt numbers, however, stress terms due to the stratified viscosity were dominant. In the inviscid limit, the corkscrew mode prevailed. The authors concluded with noting that many open questions exist pertaining to nonlinear behavior of these instabilities, as well as coupled effects of both viscosity and density stratification.

In d'Olce *et al.* (2008), two neutrally buoyant miscible fluids of different viscosities, with the lower viscosity fluid in the core, were investigated. Regardless of the core fluid radius, the flow led to two axisymmetric patterns: mushroom-like and pearl-like. Pearl shapes were associated with small core radii or low Reynolds numbers, while mushroom-like patterns were otherwise observed. More axisymmetric wave-like patterns were observed by Shum *et al.* (2010). Their work, on the other hand, involved microfluidics with droplets entrained inside the core gas.

Congruent with the findings of Selvam *et al.* (2007), Orazzo *et al.* (2014) indicated that Reynolds stresses were the dominant energy-containing terms in the annulus even at a maximum Reynolds number of 800. Their more viscous fluid was on the inside. They considered the effect of annular area on the shape of the resulting disturbances. (In our case, our annular area oscillates greatly with the wave motion.) Spiral instabilities were prevalent in the annular phase and were dominant long before axisymmetric disturbances emerged. They cite the reason that many modal analyses differ from the results of experiments is that non-modal sources of energy can prevail, citing the spiral instability as an example.

de Castro & Rodriguez (2015) considered a fully developed interfacial wave train for inclined oil-water flow, but it is unclear from their description which fluid is in the core. Perhaps the reader is to assume it is a lubricating flow (water in the annulus), since the authors cite lubrication papers. They offer modified Weber, Re, Froude numbers for correlating wave aspect ratio and dimensionless wave speed. The average wave shape experimentally obtained resembles a sea wave in the region where it is about to break. Also in 2015, Alekseenko *et al.* (2015) researched gas-centered atomization under the influence of gravity with nearly no central gas blockage by the liquid annulus. Long-wave instabilities were the most dominant. However, large-scale axisymmetric waves were a result of the coalescence of small ripples which began where the liquid and gas first meet. Despite the fact that prior work demonstrated large-scale instabilities resulted even without ripples, the authors propose that the ripples at least encourage a greater growth rate of the larger waves. Later, Tripathi *et al.*

(2017) evaluated lubricating oil-water flows, with a relatively thin annulus, demonstrating both axisymmetric and non-axisymmetric modes. By using a shear Reynolds number, they showed that the wavelength of the most dominant mode asymptotically decreases with this Reynolds number. Nearly all of the interfacial energy was stored in modes having wavelengths significantly higher than the thickness of the annular fluid.

Then, Asaduzzaman (2018) took a second look at gravity-involved vertical lubricating oil-water flow documented earlier by Li & Renardy (1999). The aim with this revisitation was to explain what happens when waves are beyond their nonlinear saturation, to understand modulation of wave amplitudes, and to offer a 3-D analysis via the commercial code Fluent. Asaduzzaman (2018) claims to use a "direct numerical simulation" (DNS), but there is no mention therein of a Kolmogorov length scale. Therefore, the present authors are skeptical if "DNS" is an appropriate designation of their work. The authors employ a volume of fluid (VOF) approach, and more specifically a piecewise linear interfacial construction (PLIC) technique, to resolve the liquid-gas interface at the sub-grid-scale (SGS). They discovered a new wave mode in which the shape shifts from saturated bamboo waves to modulated bamboo waves. They include a bifurcation diagram to characterize these wave shapes and their periods of oscillation, and they non-dimensionalize time via $tW_c/R_1$. Modulated bamboo waves were found from $0.008 < J < 0.19$, and bubbles formed and eventually leaded to broken wave crests for low $J$. They reiterated the fact that corkscrew and snake wave modes were for gravity-involved flows only.

Compressible gas flows were revisited by Li *et al.* (2019); in this case two compressible gas streams of potentially different $Ma_f$ surrounded a single liquid annular sheet. Disturbance growth rate was determined by velocity differences versus absolute velocities. They found that compressible flows had higher gas pressures near the interface, leading to higher gas densities, and thus higher aerodynamic interaction. A bifurcation emerged based on liquid Reynolds number: Above 500 (ours is 172), the instabilities were aerodynamic-driven, while below 10, they were enhanced by viscosity differences.

Another gravity-involved, incompressible, immiscible fluid flow was investigated by Sahu (2021). He considered a range of viscosities, such that viscosity ratio $m$ was above and below unity; his inverse capillary numbers $\Gamma$ ranged from 0 to 1. In his calculations, unlike ours, he had independent control over the core diameter. Corkscrew modes were dominant when the annular fluid was less viscous than the core fluid, and axisymmetric modes were less stable when the annular fluid is more viscous than the core fluid. Decreasing $a$ from 2.5 to 1.4 decreases the velocity gradient in the annular region and decreases $W_c$ to maintain the constant volumetric flow rate condition, stabilizing the flow. While increasing $\Gamma$ stabilized short-wave (high $\alpha$) modes, that same action destabilizes long-wave perturbations ($\alpha < 1$, like ours) due to Rayleigh–Plateau instability.

The gravity-influenced, incompressible, gas-driven atomization work of Alekseenko *et al.* (2015) was revisited by Cherdantsev *et al.* (2021), and it was determined that wave frequency decreased exponentially with distance from the atomizer. That was a result of the fact that wave coalescence was nearly proportional to that distance. Wave frequency at formation was nearly $3x$ that of their final values, which indicates that waves absorb two others while flowing downstream. Again, as in their earlier work, the disturbance waves were relatively small compared to the pipe diameter.

The most recent paper the present authors found on relevant work is that of Li *et al.* (2022) which involved fully developed lubricating oil-water flow. They solved the VOF problem in 2-D using OpenFOAM. Only axisymmetric waves patterns were discussed, and the work was for the purpose of pressure drop prediction.

| Descriptor | Symbol | Calculation | Value (SI, if applicable) |
|---|---|---|---|
| Core Fluid Reynolds Number (AA) | $Re_{1,AA}$ | $\frac{\rho_1 W_1 R_1}{\mu}$ | $5.56 \times 10^4$ |
| Core Fluid Reynolds Number (CL) | $Re_{1,CL}$ | $\frac{\rho_1 W_c R_1}{\mu}$ | $1.73 \times 10^5$ |
| Core Fluid Feed Mach Number | $Ma_f$ | $\frac{W_1}{a}$ | 0.225 |
| Annular Fluid Reynolds Number (AA) | $Re_{2,AA}$ | $\frac{\rho_2 W_2 R_A}{\mu}$ | 172 |
| Annular Fluid Reynolds Number (CL) | $Re_{2,CL}$ | $\frac{\rho_2 W_c R_2}{\mu}$ | $2.77 \times 10^4$ |
| Dimensionless Annulus Thickness | $\beta$ | $\frac{R_2 - R_1}{R_1}$ | 0.176 |
| Pipe: Core Radius Ratio | $a$ | $\frac{R_2}{R_1}$ | 1.18 |
| Viscosity Ratio | $m$ | $\frac{\mu_2}{\mu}$ | 5590 |
| Mobility Ratio | $M$ | $\ln(m)$ | 8.63 |
| Gas-Liquid Momentum Ratio | $\mathcal{M}$ | $\frac{\rho_1 W_1^2}{\rho_2 W_2^2}$ | 0.0792 |
| Density Ratio | $\zeta$ | $\frac{\rho_2}{\rho}$ | 761 |
| Surface Tension Parameter | $\Sigma$ | $\frac{\sigma}{\rho_1 R_1 W_c^2}$ | $6.63 \times 10^{-5}$ |
| Surface Tension Parameter | $J$ | $\frac{\sigma \rho_1 R_1}{\mu_1^2}$ | $1.98 \times 10^6$ |
| Surface Tension Parameter | $S$ | $\frac{1}{6} Re_{2,CL} \Sigma \beta^2 (2\beta + m)$ | 53.3 |
| Surface Tension Parameter | $W_{Aul}$ | $\frac{1}{3S}$ | 0.00625 |
| Pressure Gradient | $\mathcal{F}$ | $\frac{dP}{dx}$ | $2.79 \times 10^6$ |
| Capillary Length | $\kappa$ | $\sqrt{\frac{\sigma}{\mathcal{F}}}$ | $1.59 \times 10^{-4}$ |
| Capillary Number | $Ca$ | $\frac{\mu_2 * W_2}{\sigma}$ | 20.2 |
| Weber Number | $We$ | $\frac{\rho_2 c_w^2 \lambda}{\sigma}$ | $1.90 \times 10^4$ |
| Ohnesorge Number | $Oh$ | $\frac{\mu_2}{\sqrt{\rho_2 \lambda \sigma}}$ | 0.114 |
| Goucher Number | $Go$ | $\frac{R_1}{\kappa}$ | 42.4 |
| Froude Number | $Fr$ | $\frac{W_f}{\sqrt{g D_f}}$ | 201 |
| Driving Force Ratio | $K$ | $\frac{\mathcal{F} + \rho_1 g}{\mathcal{F} + \rho_2 g}$ | 1.00 |
| Inverse Capillary Number | $\Gamma$ | $\frac{\sigma}{\mu_1 W_1}$ | 35.6 |
| Dimensionless Wavenumber | $\alpha$ | $\frac{2 R_A \pi}{\lambda}$ | 0.696 |
| Dimensionless Wavelength | $\Lambda$ | $\frac{\lambda}{R_A}$ | 9.03 |
| Core Fluid "Long"/"Short" Parameter | $L_{1,LS}$ | $\frac{\lambda}{R_1 Re_{1,AA} 2\pi}$ | $4.56 \times 10^{-6}$ |
| Annular Fluid "Long"/"Short" Parameter | $L_{2,LS}$ | $\frac{\lambda}{R_A Re_{2,AA} 2}$ | $8.35 \times 10^{-3}$ |

Table 5: Derivation of dimensionless parameters based on TA data from Wilson & Strasser (2022*b*), Wilson & Strasser (2022*a*), Wilson *et al.* (2023*a*), and Wilson *et al.* (2023*b*) but newly framed in terms of CAF. "AA" implies an area-averaged velocity basis for either phase in their respective lanes, and "CL" implies centerline velocity for the core fluid to be consistent with past research by other authors. Gas density and liquid viscosity scales are based on spatiotemporal averages; the liquid viscosity scale, in particular, is volume-averaged over the region of wave formation.